\documentclass[aps,prc,superscriptaddress,twoside,twocolumn,nofootinbib,showpacs]{revtex4-2}

\usepackage{amsmath,amssymb,bm}
\usepackage{graphicx}
\usepackage{braket}
\usepackage{slashed,tensor}

\usepackage{newtxtext}
\usepackage{newtxmath}
\usepackage{mathrsfs}
\usepackage{csquotes}
\usepackage{dcolumn}

\usepackage{hyperref}
\hypersetup{colorlinks=true,citecolor=blue,linkcolor=blue,urlcolor=blue}
\newcommand{\hrefr}[2]{\href{#2}{#1}}

\begin{document}
\title{\boldmath Dynamical model of $\phi$ meson photoproduction on the nucleon and \nuclide[4]{He}}

\author{Sang-Ho Kim}
\email{shkim@pknu.ac.kr}
\affiliation{Department of Physics, Pukyong National University, Busan 48513, Korea}
\affiliation{Department of Physics and Origin of Matter and Evolution of Galaxy (OMEG) Institute,
Soongsil University, Seoul 06978, Korea}

\author{T.-S. H. Lee}
\email{tshlee@anl.gov}
\affiliation{Physics Division, Argonne National Laboratory, Argonne, Illinois 60439, USA}

\author{Seung-il Nam}
\email{sinam@pknu.ac.kr} 
\affiliation{Department of Physics, Pukyong National University, Busan 48513, Korea}
\affiliation{Asia Pacific Center for Theoretical Physics, Pohang, Gyeongbuk 37673, Korea}

\author{Yongseok Oh}
\email{yohphy@knu.ac.kr}
\affiliation{Department of Physics, Kyungpook National University, Daegu 41566, Korea}
\affiliation{Asia Pacific Center for Theoretical Physics, Pohang, Gyeongbuk 37673, Korea}

\begin{abstract}
We investigate $\phi$ meson photoproduction on the nucleon and the \nuclide[4]{He} targets within a dynamical model approach based on
a Hamiltonian which describes the production mechanisms by the Pomeron-exchange, meson-exchanges, $\phi$ radiations,
and nucleon resonance excitations mechanisms. 
The final $\phi N$ interactions are included being described by the gluon-exchange, direct $\phi N$ couplings, and the box-diagrams arising from the couplings with
$\pi N$, $\rho N$, $K\Lambda$, and $K\Sigma$ channels.
The parameters of the Hamiltonian are determined by the experimental data of $\gamma p \to \phi p$ from the CLAS Collaboration.
The resulting Hamiltonian is then used to predict the coherent $\phi$-meson production on the \nuclide[4]{He} targets by using the distorted-wave
impulse approximation.
For the proton target, the final $\phi N$ rescattering effects, as required by the unitarity condition, are found to be very weak, which supports the earlier
calculations in the literature.
For the \nuclide[4]{He} targets, the predicted differential cross sections are in good agreement with the data obtained by the LEPS Collaboration.
The role of each mechanism in this reaction is discussed and predictions for a wide range of scattering angles are presented, which can be tested in
future experiments.

\end{abstract}


\maketitle

\section{Introduction}

Photoproduction of vector mesons from nuclei has been studied to investigate nuclear shadowing and
the hadronic structure of the photon based on vector meson dominance (VMD)
hypothesis~\cite{Stodolsky67,BSYP78,Weise74,GSbook78}.
This also offers a way to study the production mechanisms from neutrons~\cite{Krusche11} and the medium modification of vector meson properties~\cite{BR91}.

Most experiments performed through photon-nucleus scatterings have been 
for semi-inclusive $\phi$ photoproduction from several nuclei, which 
allows,  with the VMD hypothesis,   
to estimate the value of the $\phi N$ total cross sections~\cite{MMMO71,IAAA04}.
Recently, exclusive $\phi$ meson photoproduction processes have been investigated at SPring-8 and Thomas Jefferson National Accelerator Facility.
The measurements for coherent and incoherent $\phi$ photoproduction from deuterium targets 
were reported in Refs.~\cite{CLAS07,LEPS07a,LEPS09b,LEPS10,CLAS10b} and, for the first time, exclusive $\phi$ photoproduction from the 
\nuclide[4]{He} targets were observed~\cite{LEPS17,Hiraiwa18}.
In the present work, we focus on the reaction of
\begin{align}
\gamma + \nuclide[4]{He} \to \phi + \nuclide[4]{He} 
\end{align}
and analyze the data reported in Ref.~\cite{LEPS17}.

Theoretical studies on coherent $\phi$ photoproduction from nuclei are rather scarce.
Most studies are for the reactions with deuteron targets~\cite{TFL02,RSS05,TK07a,SMJO10,FS13,KYD16}
and the processes with light nuclei have not been studied in detail.
The purpose of the present work is to investigate 
$\phi$ photoproduction on nuclei targets within the Hamiltonian formulation utilized by the
Argonne National Laboratory and Osaka University (ANL-Osaka) Collaboration~\cite{MSL06,KLNS19}.

In this approach we construct a model Hamiltonian with the parameters determined by the data of $\phi$ photoproduction on the nucleon
targets. 
Earlier studies of vector meson photoproduction were  mainly in the very high energy region where the Regge phenomenology 
is applicable, which led to a fairly successful Pomeron exchange model~\cite{Laget19}. 
In the near threshold energy region, however, the mechanisms arising from meson-exchanges and the excitation of nucleon resonances ($N^*$)
in the $\phi N$ channel would give non-negligible contributions as demonstrated in Refs.~\cite{TOY97,TOYM98,TLTS99,KN19,KN20}. 
In the present work, we follow the model of Refs.~\cite{KN19,KN20} for the mechanisms of $\phi$ photoproduction.

The unitarity condition requires that the $\gamma N \to \phi N$ amplitude must include the $\phi N \to \phi N$ final state interaction (FSI) as well.
As shown in the literature~\cite{KLNS19,HDHH12,ABKN11, said}, the FSI is crucial in extracting the nucleon resonances ($N^*$)  
parameters from the experimental data. 
In addition, the $\phi N \to \phi N$ reaction is essential in exploring the possible $\phi$-nucleus bound states, as predicted by lattice quantum
chromodynamics (LQCD) calculations~\cite{NPLQCD-14}.
This also accounts for the FSI in the reaction of $\phi$ meson photoproduction on \textit{nuclei\/}.
In the present work, we elaborate on the model for this reaction as well.
For this end, we construct a model for $\phi N$ interactions.
As possible sources for $\phi N$ interactions one may consider the gluon-exchange mechanism within quantum chromodynamics (QCD)
as well as the diagrams arising from non-vanishing $\phi NN$ coupling.
Another possibility is due to the decay processes of $\phi \to K \bar{K}$ and $\phi \to \pi \rho$ which then lead to the interactions through 
$\phi N \to KY, \pi N, \rho N$. 
In the leading order these interactions can generate the box-diagram mechanisms on the $\phi N$ scattering, which will be elaborated
in the present work.

With a model Hamiltonian constructed by fitting the data of $\phi$ photoproduction on the nucleon,
we will investigate its production on nuclei within the multiple scattering formulation~\cite{Feshbach92}.
By using the well-established factorization approximation, the photoproduction amplitude on a nucleus can be expressed
in terms of the $\gamma N \rightarrow \phi N$ amplitude and a nuclear form factor. 
The final $\phi$-nucleus interactions can be calculated from an optical potential which is calculated, in the leading order, from the
$\phi N \rightarrow \phi N$ amplitude and nuclear form factor. 
We will apply this approach to understand the data from the \nuclide[4]{He} targets reported by the LEPS Collaboration~\cite{LEPS17}.

This paper is organized as follows.
In Sec.~\ref{sec:gN}, we present the formulation of $\phi$ photoproduction on the nucleon.
Our dynamical model for $\phi$ photoproduction from the nucleon will be presented and discussed as well.
Section~\ref{sec:impulse} is devoted to the discussion on the Born terms of the amplitudes of $\phi$ photoproduction on the nucleon.
The FSI amplitude of the reaction is then investigated in Sec.~\ref{sec:fsi}, which completes our model for
$\phi$ photoproduction on the nucleon.
The formulation for the photoproduction on nuclei will be discussed in Sec.~\ref{sec:nuclei}, which allows us to calculate the cross sections of
$\gamma \, \nuclide[4]{He} \to \phi \, \nuclide[4]{He} $.
Our numerical results for the nucleon targets and for the \nuclide[4]{He} targets are presented in Sec.~\ref{sec:results}. 
Section~\ref{sec:summary} contains a summary and discussion.


\section{Dynamical model of $\gamma N \to \phi N$ reaction} \label{sec:gN}

\begin{figure}[t] 
\begin{center}
\includegraphics[width=\columnwidth,angle=0]{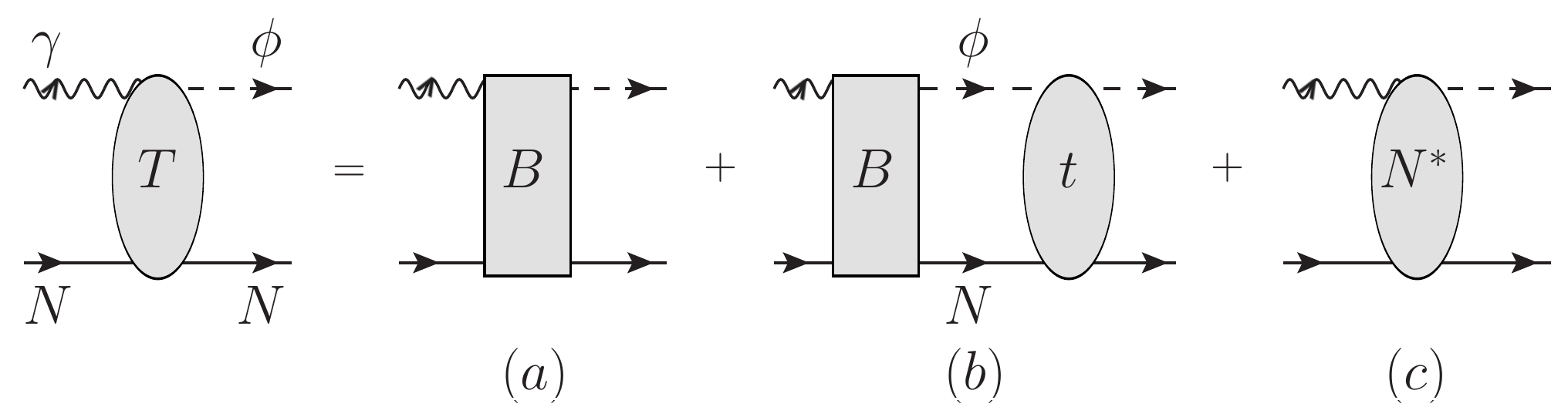}
\caption{Total amplitude of the $\gamma N \rightarrow \phi N$ reaction: $B$ is the production
amplitude, $t$ is the $\phi N$ scattering amplitude, and ${N^*}$ is the nucleon resonance contributions.
}
\label{fig:t-total}
\end{center}
\end{figure}

Following the dynamical formulation of Refs.~\cite{MSL06,KLNS19}, we first define the model Hamiltonian 
which can generate the $\gamma N \to \phi N$ reaction and the $\phi N \to \phi N$  final state interaction. 
It is also necessary to include the mechanisms induced by the $\phi$ meson decays such as
$\phi \to K\bar{K}$ and $\phi \to \rho \pi$ whose decay widths are large enough to lead to 
non-negligible coupled-channel effects arising from the one-meson-exchange mechanisms in 
$\phi N \to KY, \pi N, \rho N$ processes.
We thus consider the following form of the Hamiltonian:
\begin{align}
H =& H_0 +B_{\phi N,\gamma N} + \Gamma_{N^*,\gamma N} + \Gamma_{N^*, \phi N} 
\cr & \mbox{} 
+ \sum_{MB=K\Lambda,K\Sigma, \pi N, \rho N} \left( v_{MB,\phi N}^{} + \mbox{h.c.} \right) ,
\label{eq:model-h}
\end{align}
where $H_0$ is the free Hamiltonian of the system, $ B_{\phi N,\gamma N}$  is the Born  term
consisting of the tree diagrams for the reaction of $\gamma N \leftrightarrow \phi N$, 
and $v_{MB,\phi N}$ is the one-meson-exchange potential for ${\phi N \leftrightarrow MB}$.

\begin{figure}[t]
\centering
\includegraphics[width=\columnwidth,angle=0]{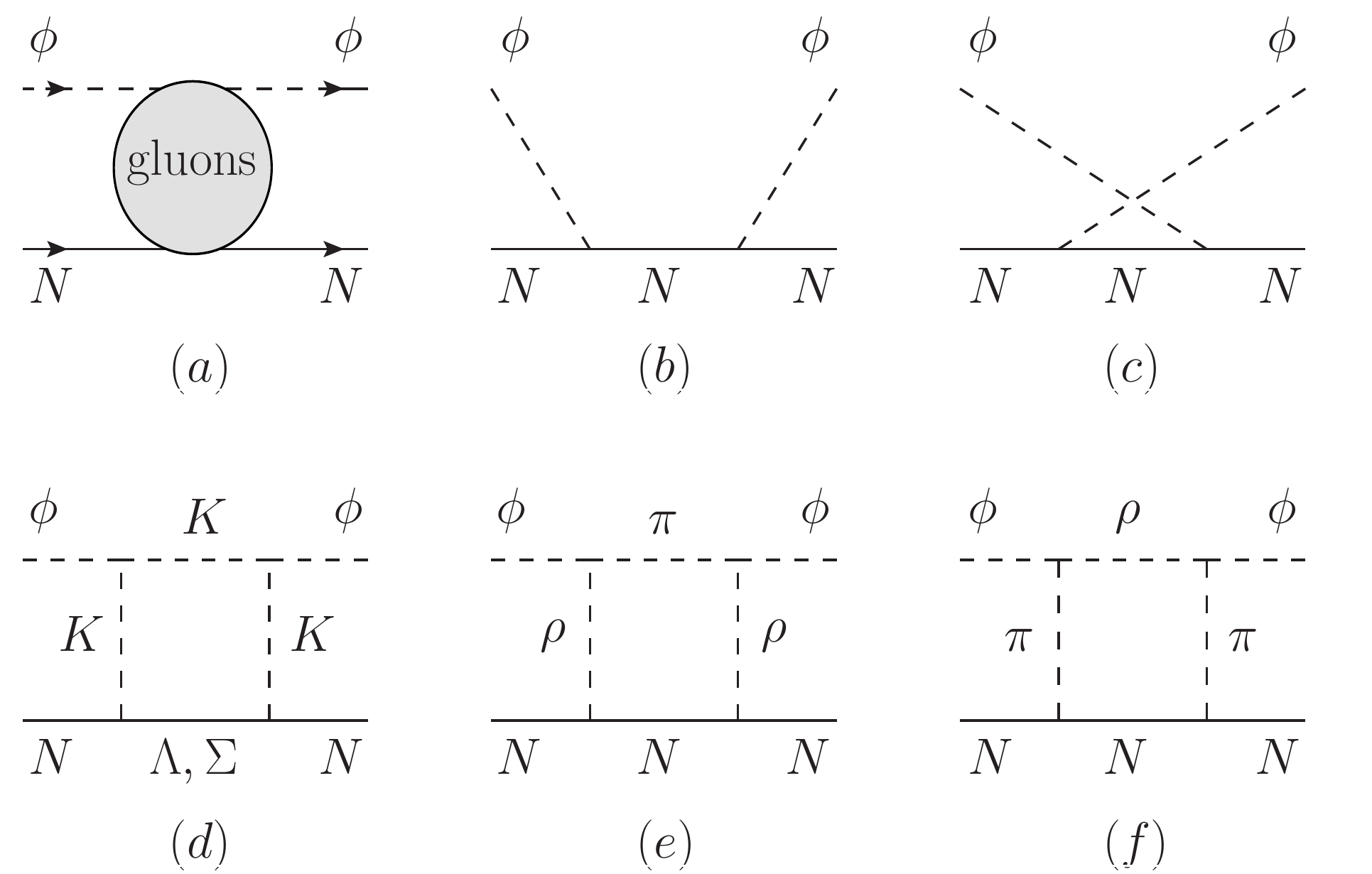}
\caption{Diagrams for the $\phi N \to \phi N$ interactions.
(a) Gluon-exchange within QCD, (b,c) mechanisms by the direct $\phi N$ coupling, and
(d,e,f) box-diagrams arising from the $\phi \to K\bar{K}$ and $\phi \to \pi \rho$ decays.}
\label{fig:phin-mech}
\end{figure}

As illustrated in Fig.~\ref{fig:t-total}, the full amplitude for the $\gamma N \to \phi N$ reaction defined 
by the above Hamiltonian can be written as
\begin{align}
T_{\phi N,\gamma N}(E) =  B_{\phi N,\gamma N} + T^{\rm FSI}_{\phi N,\gamma N}(E) 
+ T^{N^*}_{\phi N,\gamma N}(E), 
\label{eq:t-gn-phin}
\end{align}
where $T^{\rm FSI}_{\phi N,\gamma N}(E)$ and $T^{N^*}_{\phi N,\gamma N}(E)$ are, respectively, 
the amplitudes due to the $\phi N$ final state interactions and the $N^*$ contributions defined by the 
vertex functions $\Gamma_{N^*,\gamma B}$ and $\Gamma_{N^*, \phi N}$.
Explicitly, we have
\begin{align}
T^{\rm FSI}_{\phi N,\gamma N}(E) = t_{\phi N, \phi N}(E)
G_{\phi N}(E) B_{\phi N,\gamma N},
\label{eq:t-fsi}
\end{align}
where the meson-baryon propagator is
\begin{align}
G_{MB}(E)=\frac{\ket{MB} \bra{MB} }{E-H_0+i\epsilon}.
\label{eq:mb-prop}
\end{align}
The $\phi N\rightarrow \phi N$ scattering amplitude $t_{\phi N, \phi N}(E)$ in Eq.~(\ref{eq:t-fsi}) is
defined by
\begin{align}
& t_{\phi N, \phi N}(E) 
\cr & \quad
= V_{\phi N, \phi N}(E) + V_{\phi N, \phi N} \, G_{\phi N}(E) \, t_{\phi N, \phi N}(E) ,
\label{eq:lseq}
\end{align}
where the $\phi N$ potential $V_{\phi N, \phi N}(E)$ is decomposed as
\begin{align}
V_{\phi N, \phi N}(E) = v_{\phi N, \phi N}^{\rm Gluon} + v_{\phi N, \phi N}^{\rm Direct} +
v^{\rm Box}_{\phi N, \phi N}(E) ,
\end{align}
as illustrated in Fig.~\ref{fig:phin-mech}.
Here, $v_{\phi N, \phi N}^{\rm Gluon}$ is the gluon-exchange interaction [Fig.~\ref{fig:phin-mech}(a)]
and $v_{\phi N, \phi N}^{\rm Direct}$ is the direct $\phi N$ coupling term
[Figs.~\ref{fig:phin-mech}(b,c)].
The box-diagram mechanisms [Figs.~\ref{fig:phin-mech}(d-f)] are defined by 
\begin{align}
v^{\rm Box}_{\phi N, \phi N}(E) = \sum_{MB} v_{\phi N, MB}^{} \, G_{MB}^{}(E) \, v_{MB,\phi N}^{} ,
\label{eq:v-box}
\end{align}
where the intermediate meson-baryon ($MB$) states include the $K\Lambda$, $K\Sigma$, $\pi N$, $\rho N$ channels.

The $N^*$ excitation amplitude in Eq.~(\ref{eq:t-gn-phin}) is
\begin{align}
T^{N^*}_{\phi N,\gamma N}(E) = \sum_{N^*}\bar{\Gamma}^\dagger_{N^*,\phi N}
\frac{1}{E-M^{N^*}_0-\Sigma_{N^*} (E)}\bar{\Gamma}_{N^*,\gamma N},
\label{eq:t-nstar}
\end{align}
where $\bar{\Gamma}_{N^*,MB}$ are the dressed vertices, $M^{N^*}_0$ is
the bare mass of $N^*$,  and $\Sigma_{N^*}(E)$ is the self-energy of the $N^*$.
The details of these dressed $N^*$ quantities will be discussed in the next section.

By using the normalization condition $\braket{\mathbf{k} \vert \mathbf{k}'} = \delta^3 (\mathbf{k}-\mathbf{k}')$
for plane wave states~\cite{GW} and $\braket{\psi_B \vert \psi_B} = 1$ for a single particle state $\psi_B$,  
the differential cross section of $\gamma(q,\lambda_\gamma) + N(p_i^{},m_s) \to \phi({k,\lambda_\phi}) + N(p_f^{},m'_s)$ 
in the center of mass (c.m.) frame, where $\mathbf{p}_i^{} = -\mathbf{q}$ and $\mathbf{p}_f^{} = -\mathbf{k}$, 
can be written as
\begin{align}
\frac{d\sigma}{d\Omega} =& \frac{(4\pi)^2}{q^2}\rho_{\phi N}(W)\rho_{\gamma N}(W)
\frac14 \sum_{\lambda_{V},m_s'}\sum_{\lambda_\gamma,m_s}
\cr & \mbox{} \times
\left| \braket{k\lambda_\phi ; p_fm_s' \vert T_{\phi N, \gamma N}(W) \vert q\lambda_\gamma ; p_i m_s} \right|^2 ,
\end{align}
where
\begin{align}
\rho_{\phi N} =& \frac{kE_\phi(\mathbf{k})E_N(\mathbf{k})}{W},   \cr
\rho_{\gamma N} =& \frac{q^2E_N(\mathbf{q})}{W},
\end{align}
with $W = q + E_N (\mathbf{q}) = E_\phi(\mathbf{k}) + E_N(\mathbf{k})$ being the invariant mass.
Here, $\lambda_\gamma$ and $\lambda_\phi$ are the helicities of the photon and the $\phi$ meson, respectively,
and $m_s$ and $m_s'$ are the magnetic quantum numbers of the initial and final nucleons, respectively.


\begin{figure*}[t]
\begin{center}
\includegraphics[width=0.7\textwidth,angle=0]{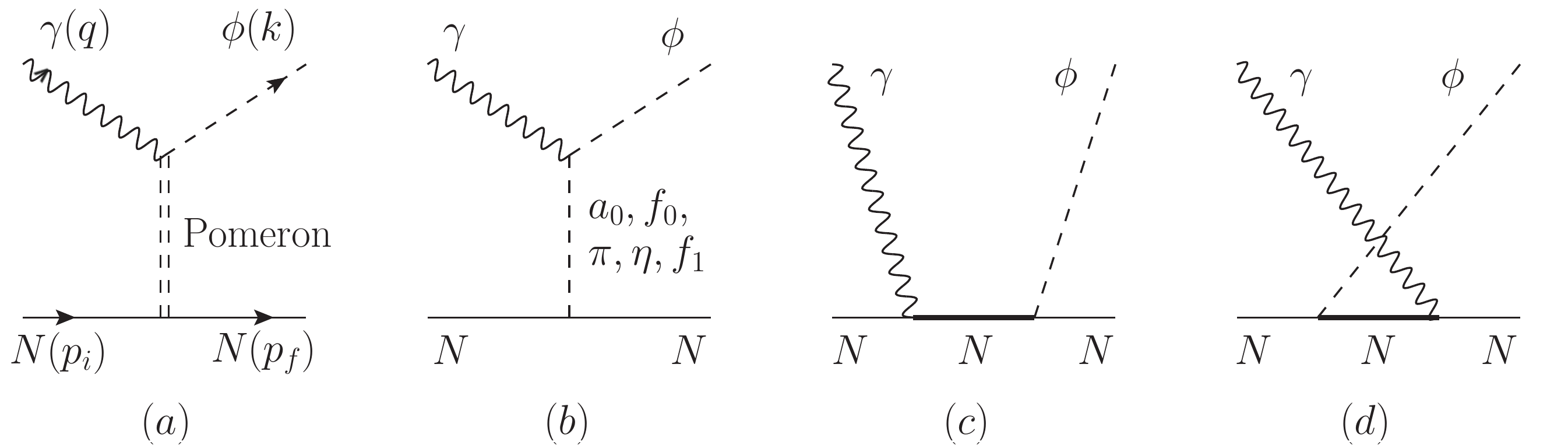}
\caption{Born terms of the $\gamma N \to \phi N$ reaction.} 
\label{fig:gn-phin}
\end{center}
\end{figure*}

\section{Born terms} \label{sec:impulse}

In the present work, we model the Born terms of the $\gamma N \to \phi N$ reaction by the diagrams
shown in Fig.~\ref{fig:gn-phin}, which defines the momenta of the involved particles as well.
Depicted in Fig.~\ref{fig:gn-phin}(a) is the Pomeron exchange mechanism and we use the parametrization of
Ref.~\cite{WL12} following the model of Donnachie and Lanshoff~\cite{DL84,DL86,DL87a,DL92}.
At low energies, however, the meson exchange mechanisms [Fig.~\ref{fig:gn-phin}(b)] and the direct
$\phi$ radiations [Fig.~\ref{fig:gn-phin}(c,d)] may give nontrivial contributions.  
In the present work, we consider these mechanisms for constructing the $\gamma N \to \phi N$ reaction amplitudes.

The amplitude for the Born term can be written as
\begin{align}
& \braket{ k\lambda_{\phi}; p_fm_s' \mid B_{\phi N, \gamma N} \mid q\lambda_\gamma; p_i m_s}
\cr \mbox{} = & \frac{1}{(2\pi)^3}\sqrt{\frac{M_N^2}{4 E_{V}(\mathbf{k}) E_N(\mathbf{p}_f^{})
\left\vert \mathbf{q} \right\vert E_N(\mathbf{p}_i^{})} }
\cr & \mbox{} \times
\left[ \bar{u}_N^{} (p_f^{},m'_s) \mathcal{M}^{\mu\nu}(k,p_f^{},q,p_i^{}) u_N^{} (p_i^{},m_s) \right]
\cr & \mbox{} \times
\epsilon^*_\nu(k,\lambda_{\phi}) \epsilon_\mu^{} (q,\lambda_\gamma) ,
\end{align}
where $\epsilon_\mu(q,\lambda_\gamma)$ is the photon polarization vector with momentum $q$ and
helicity $\lambda_\gamma$, and $\epsilon_\nu(k,\lambda_\phi)$ is that of the $\phi$ meson
with momentum $k$ and helicity $\lambda_\phi$.
The nucleon spinor of momentum $p$ and spin projection $m_s$ is represented by $u_N^{} (p,m_s)$, which is normalized as
$\bar{u}_N^{} (p,m_s) u_N^{} (p,m'_s) = \delta_{m_s,m'_s}$.
With the diagrams of Fig.~\ref{fig:gn-phin}, $\mathcal{M}^{\mu\nu}$ can be decomposed as
\begin{align}
\mathcal{M}^{\mu\nu} = \mathcal{M}_{\mathbb{P}}^{\mu\nu} + \sum_{\Phi=\pi^0,\eta} \mathcal{M}_\Phi^{\mu\nu}
+ \sum_{S=a_0,f_0} \mathcal{M}_S^{\mu\nu} +  \mathcal{M}_{f_1}^{\mu\nu}
+ \mathcal{M}_{\phi,{\rm rad}}^{\mu\nu},
\end{align}
where $\mathcal{M}_{\mathbb{P}}^{\mu\nu}$ is from the Pomeron exchange, $\mathcal{M}_\Phi^{\mu\nu}$
from pseusoscalar meson exchanges, $\mathcal{M}_S^{\mu\nu}$ from scalar meson exchanges,
$\mathcal{M}_{f_1} ^{\mu\nu}$ from $f_1(1285)$ axial-vector meson exchange,
and $\mathcal{M}_{\phi,{\rm rad}}^{\mu\nu}$ is from the direct $\phi$ radiations.
In the following subsections, each term of $\mathcal{M}^{\mu\nu}$ will be discussed in detail.


\subsection{Pomeron exchange}  \label{sec:impulse-A}

Following Refs.~\cite{OTL01,OL02,WL12}, the production amplitude of the Pomeron-exchange mechanism 
for vector meson photoproduction can be written in the form of 
\begin{align}
\mathcal{M}^{\mu\nu}_\mathbb{P}(k, p_f^{}; q, p_i^{}) = G_\mathbb{P}(s,t)
\mathcal{T}^{\mu\nu}_\mathbb{P}(k, p_f^{}; q, p_i^{}) ,
\label{eq:MP}
\end{align}
where
\begin{align}
& \mathcal{T}^{\mu\nu}_\mathbb{P}(k, p_f^{}; q, p_i^{})
\cr & \mbox{} = i \frac{12 eM_V^2}{f_V} \beta_{Q}F_V(t) \beta_{u/d} F_1(t)
\left( \slashed{q} g^{\mu\nu} - q^\nu \gamma^\mu \right) ,
\label{eq:pom-a}
\end{align}
with $t = (q - k)^2 = (p_f^{} - p_i^{})^2$. 
Here, $e$ is the unit electric charge, $M_V$ is the vector meson mass, and $f_V$ is the vector meson decay constant. 
The empirical vector meson decay constant for the $\phi$ meson is estimated as $f_V = 13.38$.%
\footnote{The value of $f_V$ is determined through the decay width of $\Gamma(V \to e^+ e^-) = 4\pi M_V \alpha_{\rm em} / (3 f_V^2)$ 
with $\alpha_{\rm em} = e^2/(4\pi)$, which leads to $f_V = 4.94$, $17.06$, $13.38$, $11.18$, and $39.68$ for $\rho^0$, $\omega$, $\phi$, $J/\psi$,
and $\Upsilon(1s)$ mesons, respectively, using the values quoted by the Particle Data Group~\cite{PDG20}.}

The coupling of the Pomeron with the quark $Q$ (or the antiquark $\bar{Q}$) in the vector meson $V$ is represented 
by $\beta_Q$ while that with the light quarks in the nucleon is given by $\beta_{u/d}$.
The Pomeron--vector-meson vertex is dressed by the form factor,
\begin{align}
F_V(t)=\frac{1}{M_V^2-t} \left( \frac{2\mu_0^2}{2\mu_0^2 + M_V^2 - t} \right) .
\label{eq:f1v}
\end{align}
By using the Pomeron-photon analogy advocated by Donnachie and Landshoff~\cite{DL84,PL97},
the form factor for the Pomeron-nucleon vertex is assumed to be the isoscalar electromagnetic
form factor of the nucleon, which can be written as
\begin{align}
F_1(t) = \frac{4M_N^2 - 2.8 t}{(4M_N^2 - t)(1-t/0.71)^2} ,
\label{eq:f1}
\end{align}
where $t$ is in unit of GeV$^2$.

The crucial ingredient of the Regge phenomenology is in the propagator $G_{\mathbb{P}}$ of
the Pomeron in Eq.~(\ref{eq:MP}), which takes the form of 
\begin{align}
G_\mathbb{P} = \left(\frac{s}{s_0}\right)^{\alpha_P^{}(t)-1} \exp\left\{ - \frac{i\pi}{2} \left[ \alpha_P^{} (t)-1 \right] \right\} ,
\label{eq:regge-g}
\end{align}
where $s=(q+p_i^{})^2=W^2$ and $ \alpha_P^{} (t) = \alpha_0^{} + \alpha'_P t$.
By fitting the cross section data of $\rho^0$, $\omega$, and $\phi$ photoproduction~\cite{OL02},
the parameters of the model have been determined to be 
\begin{eqnarray}
&& \mu_0^{} = 1.1~\mbox{GeV}^2, \quad \beta_{u/d} = 2.07~\mbox{GeV}^{-1}, \quad
\beta_{s} = 1.386~\mbox{GeV}^{-1}, 
\nonumber \\ &&
\alpha_0^{} = 1.08, \quad \alpha'_P = 1/s_0^{} = 0.25~\mbox{GeV}^{-2}.
\end{eqnarray}
For heavy quark systems, it was found~\cite{WL12}
that with the same same values of $\mu_0^2$, $\beta_{u/d}$,
and $\alpha'_P$, the $J/\psi$ and $\Upsilon(1s)$ photoproduction data could be fitted by choosing
$\beta_c = 0.323$~GeV$^{-1}$ and $\beta_b = 0.452$~GeV$^{-1}$ with $\alpha_0^{}=1.25$.
The intercept parameter $\alpha_0^{}$ of the Pomeron for heavy quarks ($c$ and $b$) production is rather different from that for the
light quarks ($u$, $d$, $s$). 
More rigorous studies are needed to understand this observation, which is, however, beyond the scope of this work.

\begin{figure}[t]
\begin{center}
\includegraphics[width=\columnwidth,angle=0]{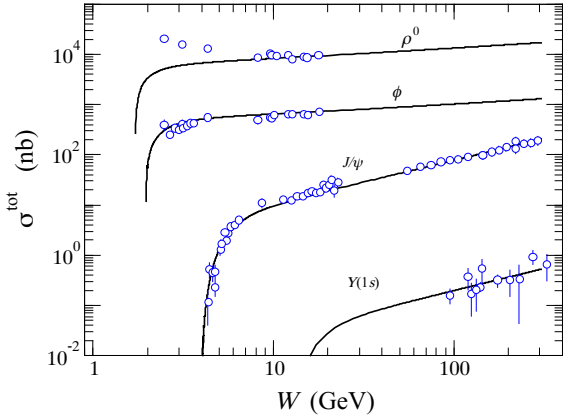}
\caption{Total cross sections for photoproduction of $\rho^0$, $\phi$, $J/\psi$, and $\Upsilon(1s)$
on the proton targets.
The data are from Ref.~\cite{HEPdata}
}
\label{fig:totcrst-all-v}
\end{center}
\end{figure}

Shown in Fig.~\ref{fig:totcrst-all-v} are the fits to the total cross section data of
$\rho^0$, $\phi$, $J/\psi$, and $\Upsilon(1s)$ mesons.
The experimental data for $\phi$, $J/\psi$, and $\Upsilon(1s)$ production processes are found to be
well described by  the Pomeron-exchange model at high energies. 
On the other hand, the $\rho^0$ and $\omega$
production data at low energies clearly need other mechanisms such as meson-exchange
mechanisms~\cite{OTL01,OL04}.

\subsection{Meson exchanges}  \label{sec:impulse-B} 

The electromagnetic interaction Lagrangians for the pseudoscalar, scalar, and $f_1(1285)$
axial-vector meson exchanges are given as
\begin{align}
\label{eq:Lag:GMPhi}
\mathscr{L}_{\gamma \Phi \phi} &= \frac{eg^{}_{\gamma \Phi \phi}}{M_\phi}
\epsilon^{\mu\nu\alpha\beta} \partial_\mu A_\nu \partial_\alpha \phi_\beta \Phi ,                    
\cr
\mathscr{L}_{\gamma S \phi} &= \frac{eg^{}_{\gamma S \phi}}{M_\phi}
F^{\mu\nu} \phi_{\mu\nu} S ,                                                                    
\cr
\mathscr{L}_{\gamma f_1 \phi} &= g_{\gamma f_1 \phi}
\epsilon^{\mu\nu\alpha\beta} \partial_\mu A_\nu
\partial^\lambda \partial_\lambda \phi_\alpha f_{1\beta} ,
\end{align}
where $\Phi$, $S$, and $f_1$ stand for the fields for the pseudoscalar, scalar, and
$f_1(1285)$ mesons, respectively.
In the present work, we consider $\Phi = \pi^0(135)$, $\eta(548)$ and $S = a_0^{}(980)$, $f_0(980)$.
The photon and $\phi$-meson field strength tensors are $F^{\mu\nu} = \partial^\mu A^\nu - \partial^\nu A^\mu$ 
and $\phi^{\mu\nu} = \partial^\mu \phi^\nu - \partial^\nu \phi^\mu$, respectively.

The coupling constants are determined by the radiative decay widths of $\phi \to \Phi \gamma$,
$\phi \to S \gamma$, and $f_1 \to \phi \gamma$, which are obtained as
\begin{align}
& \Gamma_{\phi \to \Phi \gamma} = \frac{e^2}{12\pi} \frac{q_\gamma^3}{M_\phi^2} g_{\gamma \Phi \phi}^2, 
\cr &
 \Gamma_{\phi \to S \gamma} = \frac{e^2}{3\pi} \frac{q_\gamma^3}{M_\phi^2} g_{\gamma S \phi}^2 ,
\cr & 
\Gamma_{f_1 \to \phi \gamma} = \frac{k_\gamma^3}{12\pi} \frac{M_\phi^2}{M_{f_1}^2} \left( M_{f_1}^2+ M_\phi^2 \right) g_{\gamma f_1 \phi}^2 ,
\label{eq:Wid:GMPhi}
\end{align}
where
\begin{align}
q_\gamma &= (M_\phi^2- M_{\Phi,S}^2)/(2M_\phi),   \cr
k_\gamma &= (M_{f_1}^2- M_\phi^2)/(2M_{f_1}) .
\label{eq:KinMat:RadDec}
\end{align}
Using the branching ratios data of the radiative decays~\cite{PDG20},
\begin{align}
&
\mbox{Br} (\phi \to \pi \gamma) = 1.32 \times 10^{-3},
\quad 
\mbox{Br} (\phi \to \eta \gamma) = 1.303 \times 10^{-2},
\cr &
\mbox{Br} (\phi \to a_0 \gamma) = 7.6 \times 10^{-5},
\quad 
\mbox{Br} (\phi \to f_0 \gamma) = 3.22 \times 10^{-4},
\cr &
\mbox{Br} (f_1 \to \phi \gamma) = 7.5 \times 10^{-4}, 
\label{eq:BranRatio}
\end{align}
we obtain
\begin{align}
& g_{\gamma\pi\phi} = -0.14, \qquad g_{\gamma\eta\phi}=-0.71,
\cr & 
g_{\gamma a_0 \phi} = -0.77, \qquad  g_{\gamma f_0 \phi}=-2.44 ,
\cr & 
g_{\gamma f_1 \phi} = 0.17 \mbox{ GeV}^{-2},
\label{eq:CC:GMPhi}
\end{align}
by following Ref.~\cite{TLTS99} for the phases of the coupling constants.

The strong interaction Lagrangians for describing meson exchanges are 
\begin{align}
\label{eq:Lag:MNN}
\mathscr{L}_{\Phi NN} =& -ig_{\Phi NN}^{} \, \bar N P \gamma_5 N ,      
\cr
\mathscr{L}_{S NN} =& -g_{S NN}^{} \, \bar N S N ,                    
\cr
\mathscr{L}_{f_1 NN} =& - g_{f_1 NN} \bar N
\left( \gamma_\mu - i \frac{\kappa_{f_1 NN}^{}}{2M_N} \gamma_\nu \gamma_\mu
\partial^\nu \right) f_1^\mu \gamma_5 N .
\end{align}
The strong coupling constants are obtained by using the Nijmegen potential as~\cite{SR99,RSY99}
\begin{align}
\label{eq:CC:MNN}
& g_{\pi NN}= 13.0, \qquad g_{\eta NN}= 6.34,   \cr
& g_{a_0 NN}= 4.95, \qquad g_{f_0 NN}= -0.51.
\end{align}
Following Refs.~\cite{BF95b,KMOV00}, the coupling of the $f_1$ meson is taken as
\begin{align}
|g_{f_1NN}| = 2.5 .
\label{eq:CC:f1NN}
\end{align}
We neglect the $f_1$ tensor term by setting $\kappa_{f_1 NN}^{} = 0$ in the present calculation for simplicity.

The invariant amplitudes for the pseudoscalar, scalar, and axial-vector meson exchanges read
\begin{align}
\mathcal{M}_\Phi^{\mu\nu} =& \frac{ie}{M_\phi}
\frac{g_{\gamma \Phi \phi}^{} g^{}_{\Phi NN}}{t-M_\Phi^2} \epsilon^{\mu\nu\alpha\beta} q_{\alpha}
k_{\beta} \gamma_5 ,
\cr
\mathcal{M}_S^{\mu\nu} =& \frac{e}{M_\phi}
\frac{2g_{\gamma S \phi}^{} g^{}_{S NN}}{t- M_S^2 + i\Gamma_S M_S}
(q \cdot k g^{\mu\nu} - q^\mu k^\nu) ,
\cr
\mathcal{M}_{f_1}^{\mu\nu} =& i M_\phi^2 g_{\gamma f_1 \phi} g_{f_1 NN}
\epsilon^{\mu\nu\alpha\beta}
\left( -g_{\alpha\lambda}+\frac{q_{t\alpha} q_{t\lambda}}{M_{f_1}^2} \right)
\cr &\times
\left( \gamma^\lambda + \frac{\kappa_{f_1 NN}}{2M_N} \gamma^\sigma \gamma^\lambda
q_{t\sigma} \right) \gamma_5 q_\beta P_{f_1}(t) ,
\label{eq:MesonEachAmpl}
\end{align}
where $q_t^{} = p_i^{} - p_f^{}$ and $M_H$ is the mass of hadron $H$ with $M_{a_0} = 980$~MeV and $M_{f_0} = 990$~MeV 
with $\Gamma_{a_0} \approx \Gamma_{f_0} \approx 75$~MeV~\cite{PDG20}.

In order to preserve the unitarity condition, we use the Regge prescription for $\mathcal{M}_{f_1}$.
The Regge propagator of the $f_1$ meson is given by
\begin{align}
P_{f_1}^{}(t) = \left(\frac{s}{s_{f_1}} \right)^{\alpha_{f_1}(t)-1}
\frac{\pi\alpha'_{f_1}}{\sin[\pi\alpha_{f_1}(t)]}
\frac{1}{\Gamma [\alpha_{f_1}(t)]} D_{f_1}(t) ,
\label{eq:f1:Prop}
\end{align}
where $s_{f_1} = 1$~GeV$^2$ and the $f_1$ Regge trajectory is $\alpha_{f_1} = 0.95 + 0.028 t$~\cite{KMOV00}.
The signature factor in Eq.~(\ref{eq:f1:Prop}) is of the form~\cite{KMOV00}
\begin{align}
D_{f_1}(t) = \frac{{\rm exp}[-i\pi\alpha_{f_1}(t)]-1}{2}.
\label{eq:f1:SigFac}
\end{align}

Each vertex in these amplitudes of meson exchanges is dressed by the form factor in the form of
\begin{align}
F_M (t) = \frac{\Lambda_M^4}{\Lambda_M^4 + (t - M_M^2)^2} .
\label{eq:FF:M}
\end{align}
The cutoff parameters are determined as
$(\Lambda_\Phi,\,\Lambda_S,\,\Lambda_{f_1}) =(0.25,\,1.22,\,1.50)$~GeV.

\subsection{\boldmath Direct $\phi$ meson radiations}  \label{sec:impulse-C} 

The effective Lagrangians for the direct $\phi$ radiations read 
\begin{align}
\mathscr{L}_{\gamma NN} =& - e \bar N
\left[ \gamma_\mu - \frac{\kappa_N}{2M_N} \sigma_{\mu\nu} \partial^\nu
\right] N A^\mu ,                                                      \cr
\mathscr{L}_{\phi NN} =& - g_{\phi NN} \bar N
\left[ \gamma_\mu - \frac{\kappa_{\phi NN}}{2M_N} \sigma_{\mu\nu} \partial^\nu
\right] N \phi^\mu,
\label{eq:LAG:N}
\end{align}
where $\kappa_p=1.79$.
The $\phi NN$ coupling constant is determined by using the Nijmegen potential as~\cite{SR99,RSY99}%
\footnote{
We note that smaller values of the $\phi NN$ coupling strength are obtained by
kaon loop calculations in Ref.~\cite{MMSV97}.
}
\begin{align}
\quad g_{\phi NN}=-1.47, \qquad \kappa_{\phi NN}=-1.65 .
\label{eq:CC:PhiNN}
\end{align}

The $\phi$-radiation amplitudes are then obtained as
\begin{align}
\mathcal{M}_{\phi,{\rm rad},s}^{\mu\nu} &= \frac{e g^{}_{\phi NN}}{s-M_N^2}
\left( \gamma^\nu - i\frac{\kappa_{\phi NN}}{2M_N} \sigma^{\nu\alpha} k_{\alpha}
\right)
\nonumber \\ & \quad \mbox{} \times 
 (\slashed{q}_s + M_N)
\left( \gamma^\mu + i\frac{\kappa_N}{2M_N} \sigma^{\mu\beta} q_{\beta} \right),
\cr
\mathcal{M}_{\phi,{\rm rad},u}^{\mu\nu} &= \frac{e g^{}_{\phi NN}}{u-M_N^2}
\left( \gamma^\mu + i\frac{\kappa_N}{2M_N} \sigma^{\mu\alpha} q_{\alpha} \right)
\nonumber \\ & \quad \mbox{}  \times  
(\slashed{q}_u + M_N)
\left( \gamma^\nu - i\frac{\kappa_{\phi NN}}{2M_N} \sigma^{\nu\beta} k_{\beta} \right) ,
\label{eq:Ampl:N}
\end{align}
for $s$ and $u$ channels, respectively.
The four momenta of the intermediate particles are defined as $q_s = q + p_i^{}$ and
$q_u = p_f^{} - q$.

For the form factor of the $\gamma NN$ vertex, we consider the form of Eq.~(\ref{eq:FF:M}) to have
\begin{align}
F_N (x) = \frac{\Lambda^4_N}{\Lambda^4_N+(x-M^2_N)^2} , 
\label{eq:FF:GNN}
\end{align}
for $x=(s,u)$.
Following Ref.~\cite{DW01a}, we take the common form factor as
\begin{align}
F_c (s,u) = \left[ F_N(s) + F_N(u) -F_N(s) F_N(u) \right]^2 .
\label{eq:FF:C}
\end{align}
For the $\phi NN$ vertex, we use
\begin{align}
F_N ({\mathbf k}) = \left( \frac{\Lambda_N^2}{\Lambda_N^2 + {\mathbf k}^2} \right)^2   ,
\label{eq:FF:PhiNN}
\end{align}
following the ANL-Osaka formulation, where $\mathbf{k}$ is the 3-momentum of the produced $\phi$ meson.
This choice is to ensure the convergence of the integration in calculating the FSI effect.

The final form of the $\phi$-radiation amplitude then becomes
\begin{align}
\mathcal{M}_{\phi,{\rm rad}} = (\mathcal{M}_{\phi,{\rm rad},s} + \mathcal{M}_{\phi,{\rm rad},u})
F_c(s,u) F_N({\mathbf k}) .
\label{eq:Ampl:SumN}
\end{align}
Similar amplitudes are needed to estimate the FSI effects through the $\phi p \to \phi p$ reaction as
depicted in Figs.~\ref{fig:phin-mech}(b,c).
For considering the FSI effects, we use the same form factor as given in Eqs.~(\ref{eq:FF:C})
and~(\ref{eq:FF:PhiNN}).
However, the differential cross section data of $\phi$ photoproduction in far backward
direction, $\cos\theta \leqslant -0.8$, are very limited~\cite{CLAS14}.
Since the $N$ exchange contribution rises at very large scattering angles~\cite{TLTS99},
the paucity of the data does not allow us to precisely pin down the contribution from the $N$ exchange diagrams.
In the present work, therefore, we fix its strength by $\Lambda_N = 0.98$~GeV.

\subsection{\boldmath $N^*$ excitation terms}    \label{sec:impulse-D} 

The calculation of the $N^*$ amplitude in Eq.~(\ref{eq:t-nstar}) requires a full coupled-channels calculation for the 
evaluation of the dressed vertex $\bar{\Gamma}_{N^*,\phi N}$ and the self energy $\Sigma_{N^*}(E)$.
The details can be found, for example, in Refs.~\cite{MSL06,KLNS19}.
In this exploratory study, however, we make a simplification by assuming 
$M_{N^*} - \Sigma_{N^*}(E) \sim M_{N^*} + \frac{i}{2}\Gamma^{\rm tot}(E)$, 
so that the resulting form is reduced to the usual Breit-Wigner form.
In the c.m. system, the amplitude of
$\gamma(\mathbf{q},\lambda_\gamma) + N (-\mathbf{q},\lambda_N) \to N^*(J, M_J) \to
\phi(\mathbf{k},m_\phi) + N(-\mathbf{k}, m_s')$ can then be written as~\cite{MSL06} 
\begin{align}
& \braket{ \mathbf{k}, m_\phi \, m'_s \vert T^{J} \vert \mathbf{q},\lambda_\gamma\, \lambda_N }
\cr =& 
\sum_{M_J} \braket{ \mathbf{k}, m_\phi \, m'_s \vert \bar{\Gamma}_{N^*,\phi N} \vert J M_J }
\frac{1}{W  - M_{N^*} +\frac{i}{2}\Gamma^{\rm tot}(W)}
\cr & \mbox{} \times
\braket{ JM_J \vert {\Gamma}^\dagger_{N^*,\gamma N} \vert \mathbf{q},\lambda_\gamma \, \lambda_N },
\label{eq:nstar-amp}
\end{align}
where $\lambda_\gamma$ and $\lambda_N$ are the helicities of the photon and the incoming nucleon, respectively,
and $m_\phi$ and $m_s'$ are spin projections of the $\phi$ and recoiled nucleon.
The spin and its projection of the intermediate $N^*$ are denoted by $J$ and $M_J$, respectively. 
The matrix element of the $\gamma N \to N^*$ transition is
\begin{align}
& \braket{ JM_J \vert {\Gamma}^\dagger_{N^*,\gamma N} \vert \mathbf{q},\lambda_\gamma \, \lambda_N} =
\delta_{\lambda, (\lambda_\gamma -\lambda_N)}
\cr & \qquad \mbox{}
\times \frac{1}{(2\pi)^{3/2}} \sqrt{\frac{M_N^{} q_{N^*}^{}}{E_N q}}
 A_\lambda D^J_{\lambda, M_J} (\phi_q,\theta_q,-\phi_q) ,
\end{align}
where $A_\lambda$ is the helicity amplitude of the $\gamma N \rightarrow N^*$ excitation,
$q_{N^*}^{}$ and $q$ are determined by $M_R = q_{N^*} + E_N(q_{N^*})$ and $W =q + E_N(q)$,
respectively, and
\begin{align}
D^J_{\lambda, M_J}(\phi_q,\theta_q,-\phi_q) = e^{i(\lambda-M_J)\phi}d^{J}_{\lambda,M_J}(\theta_q).
\end{align}
Here $d^{J}_{\lambda,M_J}(\theta_q)$ is the Wigner $d$-function.

\begin{figure}[t]
\centering
\includegraphics[width=\columnwidth]{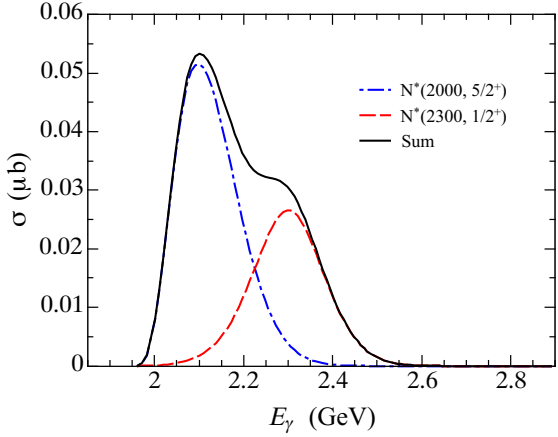}
\caption{$N^*$ contributions to the total cross section of $\gamma p \to \phi p$.}
\label{fig:nstar-comp}
\end{figure}

The matrix element of the $N^* \to \phi N$ transition is
\begin{align}
& \braket{ \mathbf{k}, m_\phi \, m'_s \vert \bar{\Gamma}_{N^*,\phi N} \vert JM_J}
\cr
=& \sum_{LS} \sum_{M_LM_S}
\braket{JM_J \vert LS M_LM_S} 
\braket{SM_S \vert 1 \textstyle\frac{1}{2}m_\phi m_{s}} 
Y_{LM_L}(\hat{\mathbf{k}}) 
\cr \times & \mbox{}
\frac{1}{(2\pi)^{3/2}}\frac{1}{\sqrt{2E_\phi(k)}}\sqrt{\frac{M_N^{}}{E_N(k)}}
\sqrt{\frac{8\pi^2M_{N^*}}{M_N k}}G^J_{LS} \left( \frac{k}{k_{N^*}} \right)^L ,
\cr
\end{align}
where $k$ and $k_{N^*}$ are determined by $W = E_\phi(k) + E_N(k)$ and
$M_{N^*} = E_\phi(k_{N^*} ) + E_N(k_{N^*})$.
The partial decay widths are defined by
\begin{align}
d\Gamma_{N^*,\gamma N} &=  (2\pi) \delta \textbf{(} M_{N^*} -e_N(q)-q \textbf{)} 
\cr & \mbox{} \times
\frac{1}{2J+1}  \sum_{M_J}\sum_{\lambda_\gamma\, \lambda_N}
\left| \braket{\mathbf{q},\lambda_\gamma\, \lambda_N \vert \bar{\Gamma}^\dagger_{N^*,\gamma N} \vert J M_J}
\right|^2 
\cr & \mbox{} \times q^2 \, dq\, d\Omega_q ,
\cr
d\Gamma_{N^*,\phi N} &= (2\pi) \delta \textbf{(} M_{N^*} -E_\phi(k)-E_N(k) \textbf{)} 
\cr & \mbox{} \times
\frac{1}{2J+1} \sum_{M_J}\sum_{m_\phi \, m_N}
\left| \braket{\mathbf{k}, m_\phi\, m'_s \vert \bar{\Gamma}_{N^*,\phi N} \vert JM_J} \right|^2 
\cr & \mbox{} \times
k^2 \, dk \, d\Omega_k .
\end{align}
Integrating over the phase space, the parameters $A_\lambda$ and $G^J_{LS}$ are found to be
related to the partial decay widths as
\begin{align}
\Gamma_{N^*,\gamma N} &= \frac{q^2_{N^*}}{4\pi} \frac{M_N}{M_{N^*}} \frac{8}{2J+1}
\left( \left\vert A_{1/2} \right\vert^2 + \left\vert A_{3/2} \right\vert^2 \right) ,
\label{eq:pwd-1} \\
{\Gamma}_{N^*,\phi N} &= \sum_{LS} \left\vert G^J_{LS} \right\vert^2 .
\label{eq:pwd-2}
\end{align}
We follow the model of Ref.~\cite{KN19} for the $\gamma N \rightarrow N^*\rightarrow \phi N$
cross sections to include  $N^*(2000,5/2^{+})$ and $N^*(2300,1/2^{+})$. 
The resonance parameters are taken similarly from Ref.~\cite{KN19}.
For $N^*(2000,5/2^{+})$, we use
\begin{align}
& \Gamma^{\rm tot} = 200~\mbox{MeV} ,
\cr & A_{1/2} = 0.031~\mbox{GeV}^{-1/2}, \quad A_{3/2} = -0.43~\mbox{GeV}^{-1/2} ,
\cr & G_{LS} = G_{2\frac{1}{2}} = 0.547~\mbox{MeV} ,
\end{align}
and the parameters of $N^*(2300,1/2^{+})$ are 
\begin{align}
& \Gamma^{\rm tot} = 300~\mbox{MeV} ,
\cr & A_{1/2} = 0.031~\mbox{GeV}^{-1/2}, \quad A_{1/2} = 0.0 ,
\cr & G_{LS} = G_{0\frac{1}{2}} = {2.80}~\mbox{MeV} .
\end{align}
We also employ the Gaussian form factor~\cite{KN19}
\begin{align}
F_{N^*} (s)= \exp\left[ \frac{-(s-M^2_{N^*})^2}{\Lambda_{N^*}^4} \right] ,
\end{align}
with $\Lambda_{N^*} = 1.0$~GeV.
Our results are shown in Fig.~\ref{fig:nstar-comp}, which is similar to Fig.~3(b) of Ref.~\cite{KN19}.

\section{The FSI amplitude} \label{sec:fsi}

The amplitude with final state interactions is defined in Eqs.~(\ref{eq:t-fsi})-(\ref{eq:v-box}). 
In the present work, we focus on examining the relative importance among
the gluon exchange term $v^{\rm Gluon}_{\phi N, \phi N}$, direct $\phi N$ coupling term
$v^{\rm Direct}_{\phi N, \phi N}$, and box-diagrams term $v^{\rm Box}_{\phi N, \phi N}$.
This can be done by keeping the leading term in Eq.~(\ref{eq:lseq}) with taking the approximation 
that $t_{\phi N, \phi N}(E) \sim V_{\phi N, \phi N}(E)$ in evaluating
the FSI amplitude in Eq.~(\ref{eq:t-fsi}).
In the c.m. frame, the amplitude  $T^{\rm FSI}_{\phi N, \gamma N}(E)$ of Eq.~(\ref{eq:t-fsi}) for the reaction of
$\gamma (\mathbf{q}) + N(-\mathbf{q}) \to \phi(\mathbf{k}) + N(-\mathbf{k})$ can then be obtained as 
\begin{align}
& \braket{\mathbf{k} \vert T^{\rm FSI}_{\phi N, \gamma N}(E) \vert \mathbf{q} }
=\int d\mathbf{k}' \braket{\mathbf{k} \vert V_{\phi N, \phi N}(E) \vert \mathbf{k}'}
\cr & \mbox{} \times 
\frac{1}{E-E_\phi (k')-E_N(k')+i\epsilon}
\braket{ \mathbf{k}' \vert B_{\phi N, \gamma N}(E) \vert \mathbf{q}}.
\label{eq:fsi-born}
\end{align}

The main task is then to evaluate the matrix elements of the  potential of $V_{\phi N,\phi N}(E)$ for
$\phi (k') + N (p') \rightarrow \phi (k) + N(p)$, which can be written as
\begin{align}
& \braket{ k \lambda_{\phi}; p m_s \vert V_{\phi N, \phi N} \vert k'\lambda'_\phi; p' m'_s}
\cr  = & \mbox{}
 \frac{1}{(2\pi)^3}\sqrt{\frac{ M_N^2 }{4 E_\phi (\mathbf{k}) E_N(\mathbf{p})
E_\phi(\mathbf{k}') E_N(\mathbf{p}^{\,\,'}) }} 
\cr & \mbox{} \times
\mathcal{V}(k\lambda_\phi, p m_s; k' \lambda'_\phi ,p' m'_s),
\end{align}
where $E_\phi$ and $E_N$ are the energies of the $\phi$ meson
and the nucleon, respectively, and
\begin{align}
\mathcal{V}=\mathcal{V}_{\rm Gluon} + \mathcal{V}_{\rm Direct} +
\sum_{MB=K\Lambda,K\Sigma,\pi N,\rho N}\mathcal{V}_{MB}(E) .
\end{align}
Here, $\mathcal{V}_{\rm Gluon}$ [Fig.~\ref{fig:phin-mech}(a)] and $\mathcal{V}_{\rm Direct}$
[Figs.~\ref{fig:phin-mech}(b,c)] are from the gluon-exchange interaction and the direct $\phi N$ coupling term, 
respectively, and $\mathcal{V}_{MB}(E)$ [Figs.~\ref{fig:phin-mech}(d-f)] includes the box-diagram mechanisms
defined by Eq.~(\ref{eq:v-box}).
In the following subsections, we elaborate on calculating these $\phi N$ potentials
from the interaction Lagrangians by using the unitary transformation method of the ANL-Osaka
formulation~\cite{KLNS19}.

\subsection{Gluon exchange interaction}

Because of the OZI rule, the $\phi N$ interaction is expected to be governed by gluon exchanges.
However, since there exists no LQCD calculations for the $\phi N$ potential, we use the form suggested 
by the recent analysis of Ref.~\cite{KS10b} for the charmonium-nucleon potential.
It was found that  that the calculated charmonium-nucleon potential is approximately of the Yukawa form
which has also been assumed in phenomenological studies~\cite{GLM00} of the $\phi N$ interactions. 
We, therefore, take the form of
\begin{align}
\mathcal{V}_{\rm gluon}=-v_0^{} \frac{e^{-\alpha r}}{r} .
\label{eq:yukawa}
\end{align}
For the charmonium-nucleon system, the LQCD data of Ref.~\cite{KS10b} can be approximated by 
the above form with $v_0=0.06$ and $\alpha = 0.3$~GeV.
Since the $\phi N$ potential is expected to have different range and the strength, we consider the range 
of parameters as $0.1 < v_0^{} < 1.0$ and $0.3 < \alpha < 0.6 $~GeV.
As will be discussed in Sec.~\ref{sec:results}, the best fit to the $\phi$ photoproduction
data was obtained by setting $v_0 =0.2$ and $\alpha = 0.5$~GeV.

The potential of Eq.~(\ref{eq:yukawa}) can be obtained by taking the nonrelativistic limit of
the scalar meson exchange amplitude calculated from the Lagrangian,
\begin{align}
\mathscr{L}_{\sigma} = V_0 \left( \bar{\psi}_N\psi_N \Phi_\sigma + \phi^\mu\phi_\mu\Phi_\sigma \right) ,
\label{eq:lag-s}
\end{align}
where $\Phi_\sigma$ is a scalar field with mass $\alpha$ in Eq.~(\ref{eq:yukawa}). 
By using the unitary transformation method used by the ANL-Osaka formulation,
the scalar-meson exchange matrix element derived from Eq.~(\ref{eq:lag-s}) is in the form of
\begin{align}
&\mathcal{V}_{\rm gluon}(k\lambda_\phi,p m_s;k'\lambda'_\phi,p'm'_s) = 
\frac{V_0}{(p-p')^2 - \alpha^2}
\cr &\, \mbox{} \times
[ \bar{u}_N(p,m_s) u_N(p',m'_s)]
[ \epsilon^*_\mu(k,\lambda_\phi) \epsilon^\mu (k',\lambda'_\phi)],
\label{eq:yukawa-rel}
\end{align}
where $V_0 = -8 v_0^{} \pi M_\phi$
and ${(p - p')}
= \textbf{(} E_N(p)-E_N(p'),\,\mathbf{p}-\mathbf{p}' \textbf{)}$.
We will use this form in our calculations.

\subsection{\boldmath Direct $\phi N$ coupling term}

The form of the direct $\phi N$ coupling amplitudes is the same as given in Eqs.~(\ref{eq:Ampl:N})
and (\ref{eq:Ampl:SumN}) after the replacement of $e$ by $g_{\phi NN}$
and $\kappa_N^{}$ by $\kappa_{\phi  N N}^{}$.
We also use the same form of the form factors as given in Eqs.~(\ref{eq:FF:GNN})-(\ref{eq:FF:PhiNN}) and
the same cutoff parameter, i.e., $\Lambda_N = 0.98$~GeV.

\subsection{Box-diagram mechanisms}

To calculate the box-diagrams depicted in Figs.~\ref{fig:phin-mech}(d-f),
the transition potentials for $\phi N \to K\Lambda,\,K\Sigma,\,\pi N,\,\rho N$ are needed, 
which can be constructed by the interaction Lagrangians given below.
The $\phi N \rightarrow K Y(Y = \Lambda,\Sigma)$ processes are described by
\begin{align}
\mathscr{L}_{\phi K{K}} &= i g_{\phi KK}^{}
\left( K^- \partial_\mu {K^+} - \partial_\mu K^- K^+ \right) \phi^\mu ,
\label{eq:Lag-a} \\
\mathscr{L}_{KNY} &= -\frac{f_{KNY}}{M_K}
\bar Y \gamma_\mu\gamma_5\,N \partial^\mu K + \mbox{h.c.},
\label{eq:Lag-b}
\end{align}
where the coupling constant $g_{\phi KK} = 4.48$ is determined from the experimental data for the decay width, 
$\Gamma_{\phi \to K^+ K^-} = 2.09$~MeV, where
\begin{align}
\Gamma_{\phi \to K^+ K^-} = \frac{g_{\phi KK}^2 p_K^3}{6 \pi M_\phi^2},
\label{eq:Kinem:phiKk}
\end{align}
with $p_K^{} = \sqrt{M_\phi^2 - 4M_K^2}/2$.

For the $\phi N \rightarrow \pi N$ and $\phi N \to \rho N$ processes, we use
\begin{align}
\mathscr{L}_{\phi\rho\pi} =& -\frac{g_{\phi\rho\pi}}{M_\phi}
\epsilon^{\mu\nu\alpha\beta}
(\partial_\mu \bm{\rho}_\nu) \cdot (\partial_\alpha \bm{\pi}) \phi_\beta ,
\label{eq:Lag-1} \\
\mathscr{L}_{\rho NN} =& g_{\rho NN}\bar{N} 
\left(  \bm{\rho}_\mu \gamma^\mu - \frac{\kappa_{\rho NN}}{2M_N} \sigma^{\mu\nu} \partial_\nu \bm{\rho}_\mu \right) \cdot \bm{\tau} N ,
\label{eq:Lag-2} \\
\mathscr{L}_{\pi NN} =& -\frac{f_{\pi NN}}{M_\pi}\bar{N}\gamma_\mu\gamma_5 \bm{\tau} \cdot
(\partial^\mu \bm{\pi}) N ,
\label{eq:Lag-3}
\end{align}
where $g_{\phi\rho\pi} = 1.22$ is determined by the decay width of $\phi \rightarrow \rho\pi$.
The $\rho NN$ coupling constant is determined by using the Nijmegen potential~\cite{SR99,RSY99},
\begin{align}
g_{\rho NN}=2.97, \quad \kappa_{\rho NN}=4.22.
\label{eq:CC:RhoNN}
\end{align}
The couplings of pseudoscalar mesons and baryons are determined by the SU(3) flavor symmetry relations,
\begin{align}
\frac{f_{KN\Lambda}}{M_K} &= \frac{-3+2\alpha}{\sqrt3} \frac{f_{\pi NN}}{M_\pi} ,
\cr
\frac{f_{KN\Sigma}}{M_K} &= (-1 + 2\alpha) \frac{f_{\pi NN}}{M_\pi}.
\label{eq:Coupl:PSBB}
\end{align}
With $\alpha = 0.635$ and $f_{\pi NN} / \sqrt{4\pi} = \sqrt{0.08}$, we get $f_{KN\Lambda} = -3.46$ and $f_{KN\Sigma} = 0.92$.

For the reaction of $\phi(k') + N(p') \to M(k) + B(p)$, the amplitudes of $K$-, $\rho$-, and $\pi$-exchanges derived by using the 
unitary transformation method from the above Lagrangians can be expressed as
\begin{align}
& I_{MB,\phi N} (k, p m_s ; k'\lambda_\phi', p' m_s')
\cr & = \bar{u}_B(p,m_s) I_{MB,\phi N}^\mu u_N(p',m_s') \epsilon_\mu (k',\lambda_\phi') ,
\label{eq:Ampl:PhiN_MN}
\end{align}
where
\begin{align}
& I_{KY,\phi N}^\mu =
i \frac{g_{\phi KK} f_{KNY}}{M_K} \frac{1}{q^2 - M^2_K }\slashed{q} \gamma_5 (k - q)^\mu ,
\label{eq:BoxAmpl-a} \\
& I_{\pi N, \phi N}^\mu  =
-\frac{g_{\phi\rho\pi}^{} g_{\rho NN}^{}}{M_\phi} \frac{1}{q^2-M^2_\rho} \epsilon^{\mu\nu\alpha\beta} k_\nu q_\alpha
\cr & \hspace{4.3em} \mbox{} \times
\left[ \gamma_\beta - \frac{\kappa_{\rho NN}}{4M_N}
\left(\gamma_\beta \slashed{q} - \slashed{q} \gamma_\beta \right) \right] ,
\label{eq:BoxAmpl-b} \\
& I_{\rho N,\phi N}^\mu = \sum_{\lambda_\rho}
-i \frac{g_{\phi\rho\pi}}{M_\phi} \frac{f_{\pi NN}}{M_\pi} \frac{1}{q^2-M^2_\pi} \epsilon^{\mu\nu\alpha\beta}
k_\nu q_\alpha
\cr & \hspace{4.3em} \mbox{} \times
\epsilon^*_\beta (k,\lambda_\rho) \slashed{q} \gamma_5 .
\label{eq:BoxAmpl-c}
\end{align}
Here, $q = p' - p = \textbf{(} E_N(p') - E_{B} (p),\,\mathbf{p}' - \mathbf{p} \textbf{)}$, $B=Y$ or $N$, and
the polarization vector $\epsilon_\beta (k,\lambda_\rho)$ of the $\rho$ meson with momentum $k$ and helicity $\lambda_\rho$ is
introduced.

We consider the following form factor for the $\phi M$ vertex
\begin{align}
F_{\rm Box} (\mathbf{k}_{MB}) =
\left( \frac{\Lambda_{MB}^2}{\Lambda_{MB}^2 +\mathbf{k}_{MB}^2} \right)^2 ,
\end{align}
and
\begin{align}
F_{\rm Box}' (\mathbf{k}_{MB}') =
\left( \frac{\Lambda_{MB}'^2}{\Lambda_{MB}'^{2} + \mathbf{k}_{MB}'^{2} } \right)^2 ,
\end{align}
for the $NB$ vertex, where
\begin{align}
\mathbf{k}_{MB}^2 &= \sum_{i=1,3}\left( \frac{\mathbf{k}_i q_0^{} -\mathbf{q}_i^{} k_0^{}}{q_0^{} + k_0^{}} \right)^2 , \cr
\mathbf{k}_{MB}^{'2} &= \mathbf{q}^2 = (\mathbf{p' - p})^2 .
\end{align}
We choose the cutoff parameters to be $650$-$1200$~MeV which are in the range of
the meson-exchange amplitude determined in the ANL-Osaka analysis of the $\pi N$ and $\gamma N$
reactions. 
Explicitly, their values are chosen as
\begin{align}
&\Lambda_{K\Lambda,K\Sigma,\pi N, \rho N} = 860~\mbox{MeV} , \quad
\Lambda_{K\Lambda,K\Sigma}' = 1200~\mbox{MeV},  \cr
& \Lambda_{\pi N}' = 920~\mbox{MeV}, \qquad \Lambda_{\rho N}' = 656~\mbox{MeV} .
\label{eq:FormFactors}
\end{align}

In the c.m. frame, in which the scattering cross sections are evaluated,
the matrix element of the box-diagram mechanism is calculated from
\begin{align}
& \mathcal{V}_{\rm MB}(k\lambda_\phi,p m_s;k'\lambda'_\phi,p'm'_s)
\cr &= \sum_{MB=K\Lambda,\pi N,\rho N}
\int d\bm{\kappa} \, I_{\phi N, MB}(k\lambda_\phi,p m_s;p_M,p_Bm_B)
\cr & \quad \mbox{}
\times\frac{1}{(2\pi)^3}\frac{1}{2E_M(\bm{\kappa})}\frac{M_B}{E_B(\bm{\kappa})}
\frac{1}{W-E_M(\bm{\kappa})-E_B(\bm{\kappa})+i\epsilon} 
\cr & \quad \mbox{}
\times I_{MB,\phi N}(p_M, p_Bm_B;k'\lambda'_\phi,p' m'_s) ,
\end{align}
where
$\textbf{p} = -\textbf{k}$, $\textbf{p}' = -\textbf{k}'$, $p_M=(E_M,\bm{\kappa})$, and 
$p_B^{} = (E_B,-\bm{\kappa})$.

\section{Formulation for $A(\gamma,\phi)A$ reaction} \label{sec:nuclei}

The differential cross section of coherent photoproduction of a vector meson ($V$) on a nuclear
target ($A$) with $A$ nucleons, $\gamma(q) + A(P_i) \rightarrow V(k) + A(P_f)$, are obtained as
\begin{align}
\frac{d\sigma}{dt}=\frac{\pi}{\lvert \mathbf{q} \rvert \lvert \mathbf{k} \rvert}
\frac{d\sigma}{d\Omega_{\rm Lab}} , 
\end{align}
where the differential cross section in the laboratory (Lab) frame ($\mathbf{P}_i=0$) is
\begin{align}
\frac{d\sigma}{d\Omega_{\rm Lab}} =&
\frac{(2\pi)^4 \left\vert \textbf{k} \right\vert^2 E_{V}(\textbf{k}) E_A(\textbf{q}-\textbf{k})}
{\left\vert E_A(\textbf{q} - \textbf{k}) \vert \textbf{k} \vert + E_{V}(\mathbf{k})
( \vert \mathbf{k} \vert - \vert \bm{q} \vert \cos\theta_{\rm Lab}) \right\vert}
\cr & \mbox{} \times
\frac{1}{2(2J+1)}
\cr & \mbox{} \times
\left\lvert \braket{k\lambda_V, \Phi_{P_f M_{J_f}} \lvert T(E) \rvert q\lambda_\gamma, P_i
\Phi_{P_i,M_{J_i}}} \right\rvert^2 ,
\end{align}
where $\cos\theta_{\rm Lab}=\hat{\mathbf{q}} \cdot \hat{\mathbf{k}}$.
Within the distorted-wave impulse approximation of multiple scattering theory~\cite{Feshbach92},
the scattering $T$-matrix defined by the Hamiltonian of Eq.~(\ref{eq:model-h}) can be written as
\begin{eqnarray}
T(E)= T^{\rm IMP}(E)+T^{\rm FSI}(E),
\end{eqnarray}
where
\begin{eqnarray}
T^{\rm IMP} &=& \sum_{i=1,A} \left[ B_{\phi N_i,\gamma N_i} + T^{N^*}_{\phi N_i,\gamma N_i} \right],
\nonumber \\
T^{\rm FSI}(E) &=& T_{\phi A, \phi A}(E) \frac{1}{E-H_0} T^{\rm IMP} .
\end{eqnarray}
The impulse term $T^{\rm IMP}$ is the term that the $\phi$ meson is produced from a single nucleon in
the nucleus, and $T^{\rm FSI}$ is the effect due to the scattering of the outgoing $\phi$ with the recoiled nucleus.
The $\phi A \rightarrow \phi A$ scattering $T$-matrix is defined by
\begin{eqnarray}
T_{\phi A,\phi A}(E) &=& U_{\phi A,\phi A}(E)
\cr && \mbox{} +U_{\phi A,\phi A}(E)
\frac{1}{E-H_0+i\epsilon}T_{\phi A,\phi A}(E),
\end{eqnarray}
where $U_{\phi A,\phi A}(E)$ is the $\phi A$ potential.

Within the multiple scattering theory~\cite{Feshbach92}, one can define the $\phi A$ potential
in terms of the $\phi N$ scattering amplitude. 
In the first order, we have 
\begin{align}
U_{\phi A,\phi A}(E)=\sum_{i=1,A}t_{\phi N_i, \phi N_i}(\omega),
\end{align}
where $t_{\phi N_i, \phi N_i}(\omega)$ was defined in Eq.~(\ref{eq:lseq}).
We take the widely used factorization approximation~\cite{Feshbach92} to evaluate the
matrix element of $U_{\phi A,\phi A}(E)$.
In the c.m. frame, for the reaction of $\phi(\bm{\kappa})+A(-\bm{\kappa}) \to \phi(\bm{\kappa}') + A(-\bm{\kappa}')$,
we have
\begin{align}
U_{\phi A,\phi A}(\bm{\kappa},\bm{\kappa}',E) 
&= \braket{ \bm{\kappa}' \vert U_{\phi A, \phi A} \vert \bm{\kappa}}
\cr &\hspace{-5em} = A \braket{ \bm{\kappa}', (\mathbf{p}_0+\mathbf{q}) \vert
t_{\phi N,\phi N}(\omega_0) \vert \bm{\kappa},\mathbf{p}_0} F(q), 
\label{eq:1st-opt}
\end{align}
where $\mathbf{q}=\bm{\kappa}-\bm{\kappa}'$, $\mathbf{p}_0 = -\bm{\kappa}/A$,
$\omega_0=E_\phi(\kappa_0^{})+E_N(\kappa_0^{}/A)$ with $\kappa_0^{}$ defined by
$E=E_\phi(\kappa_0^{})+E_A(\kappa_0^{})$, and 
\begin{align}
F(q)=\int d\mathbf{r} \, e^{i\mathbf{q}\cdot\mathbf{r}}\rho(\mathbf{r}).
\end{align}
Here $\rho(\mathbf{r})$ is the nuclear density normalized as
$\int d\mathbf{r}\, \rho(\mathbf{r})=1$.

\subsection{Cross sections from the impulse term}

By using the factorization approximation within the multiple scattering
formulation~\cite{Feshbach92}, the contribution from the impulse term $T^{\rm IMP}$ for spin
$J=0$ nuclei can be written as
\begin{align}
\frac{d\sigma^{\rm IMP}}{d\Omega_{\rm Lab}} = & 
\frac{(2\pi)^4 \vert \mathbf{k} \vert^2 E_{V}(\mathbf{k}) E_A (\mathbf{q} - \mathbf{k})}
{ \left\vert E_A(\mathbf{q} - \mathbf{k}) \vert \mathbf{k} \vert + E_{V}(\mathbf{k})
( \vert \mathbf{k} \vert - \vert \mathbf{q} \vert \cos\theta_{\rm Lab} ) \right\vert }
\cr & \mbox{} \times
\left\vert AF_T(t) \bar{t}(\mathbf{k},\mathbf{q}) \right\vert^2,
\label{eq:dsdt}
\end{align}
where $t=(q-k)^2$ and $\bar{t}(\mathbf{k},\mathbf{q})$ is the spin-averaged $\gamma N \to V N$
amplitude defined by
\begin{align}
\left\vert \bar{t}(\mathbf{k},\mathbf{q}) \right\vert^2 =  
\frac{1}{4} \sum_{m_s,\lambda_\gamma} \sum_{m_s',\lambda_{V}}
\left\vert \braket{k\lambda_{V}; p_f^{} m_s' \vert T_{VN,\gamma N} \vert q\lambda_\gamma;  p_i^{} m_s}
\right\vert^2 .
\end{align}
Here, $\braket{k\lambda_{V}; p_f^{} m_s' \vert T_{VN,\gamma N} \vert q\lambda_\gamma; p_i^{} m_s}$
is the matrix element of the $\gamma(q)+N(p_i)\rightarrow V(k)+N(p_f)$ process as given in
the previous section.
The initial nucleon momentum $p_i^{}$ in the initial target state is usually chosen as $p_i=(M_N,\bm{0})$ 
by the frozen nucleon approximation and the final nucleon momentum $p_f^{} = (p^0_f,\mathbf{p}_f^{})$ is
set as
\begin{align}
\mathbf{p}_f^{} &= \mathbf{p}_i^{} + ( \mathbf{q} - \mathbf{k} ),
\cr p^0_f &= E_N(\mathbf{p}_f^{}) = \sqrt{\mathbf{p}_f^{2} + M^2_N} .
\end{align}
We follow the standard Hamiltonian formulation within which the $\gamma + N \to V + N$
process in nuclei can be off-energy shell, i.e.,
$q^0 + E_N(\mathbf{p}_i^{}) \neq E_{V} (\mathbf{k}) + E_N(\mathbf{p}_f^{})$.

The factor $F_T(t)$ in Eq.~(\ref{eq:dsdt}) is a nuclear form factor which is probed by the
gluon-exchange mechanism.
Within the Pomeron-exchange model of Donnachie and Lanshoff~\cite{DL84}, as used in
Refs.~\cite{OL02,WL12}, the same form factor is used in usual hadron-nuclear reactions and is
defined as
\begin{align}
F_T(t)= \braket{ \Psi_T \vert \sum_{i=1,A}e^{i \bm{\kappa} \cdot \bm{r}^{}_i} \vert \Psi_T},
\label{eq:ft}
\end{align}
where $\ket{\Psi_T}$, normalized as $\braket{\Psi_T \vert \Psi_T} = 1$, is the nuclear
ground state in the nuclear c.m. frame, and $\bm{\kappa}$ is related to $t$ by
\begin{align}
-t = \bm{\kappa}^{2}-\omega^2,
\end{align}
with
\begin{align}
\omega =\sqrt{\bm{\kappa}^{\,\,2} + M^2_T} - M_T^{} .
\end{align}
Here $M_T^{}$ is the mass of the target nucleus $T$.
Clearly, $F_T(t)$ is related to the nuclear charge form factor $F_c(q^2)$
(with no exchange current contribution) by
\begin{align}
F_c(q^2)=F_N(q^2) F_T(q^2=t) ,
\end{align}
where $F_N(q^2)$ is the nucleon charge form factor.

\subsection{Cross sections including FSI}

Including the FSI term, the differential cross sections are calculated as
\begin{align}
\frac{d\sigma^{\rm }}{d\Omega_{\rm Lab}} = &
\frac{(2\pi)^4 \vert \mathbf{k} \vert^2 E_{V}(\mathbf{k}) E_A (\mathbf{q} - \mathbf{k})}
{ \left\vert E_A(\mathbf{q} - \mathbf{k}) \vert \mathbf{k} \vert + E_{V}(\mathbf{k})
( \vert \mathbf{k} \vert - \vert \mathbf{q} \vert \cos\theta_{\rm Lab} ) \right\vert }
\cr & \mbox{} \times
\left\vert AF_T(t) \bar{t}(\mathbf{k},\mathbf{q})+ T^{\rm FSI}(\mathbf{k},\mathbf{q},E)\right\vert^2,
\label{eq:dsdt-imp-fsi}
\end{align}
where
\begin{eqnarray}
T^{FSI}(\mathbf{k},\mathbf{q},E)&=&\int d\mathbf{k}'\, 
T_{\phi A,\phi A}(\mathbf{k},\mathbf{k}',E)
\nonumber \\
&&\times \frac{ AF(t')\bar{t}(\mathbf{k}',\mathbf{q}) }{E-E_V(\mathbf{k}')-E_A(\mathbf{q}-\mathbf{k}')+i\epsilon} 
 \label{eq:fsi}
\end{eqnarray}
with $t'=(q-k')^2$.
It is most convenient to evaluate $t_{\phi A,\phi A}(\mathbf{k},\mathbf{k}',E)$
in the $\phi A$ c.m. system, which leads to
\begin{eqnarray}
T_{\phi A,\phi A}(\mathbf{k},\mathbf{k}',E) &=& 
\sqrt{\frac{E_\phi(\bm{\kappa}) E_A(-\bm{\kappa})} {E_\phi(\mathbf{k})E_A(\mathbf{k}+\mathbf{q})}} 
T_{\phi A,\phi A}(\bm{\kappa},\bm{\kappa}',E) 
\nonumber \\ && \mbox{} \times 
\sqrt{\frac{E_\phi(\bm{\kappa}')E_A(-\bm{\kappa}')}
{E_\phi(\mathbf{k}')E_A(\mathbf{k}'+\mathbf{q})}}.
\end{eqnarray}
Here, $T_{\phi A,\phi A}(\bm{\kappa},\bm{\kappa}',E)$ is calculated as
\begin{eqnarray}
&& T_{\phi A,\phi A}(\bm{\kappa},\bm{\kappa}',E)
\nonumber \\ &=&
U_{\phi A,\phi A}(\bm{\kappa},\bm{\kappa}',E)
\nonumber \\ && \mbox{} 
+ \int d \bm{\kappa}''
U_{\phi A,\phi A}(\bm{\kappa},\bm{\kappa}'',E)
\frac{1}{E-E_V(\bm{\kappa}'')-E_A(\bm{\kappa}'')+i\epsilon} 
\nonumber \\ && \mbox{} \times
T_{\phi A,\phi A}(\bm{\kappa}'',\bm{\kappa}',E) .
\label{eq:ls-cm}
\end{eqnarray}

\begin{figure*}[t]
\centering
\includegraphics[width=\textwidth]{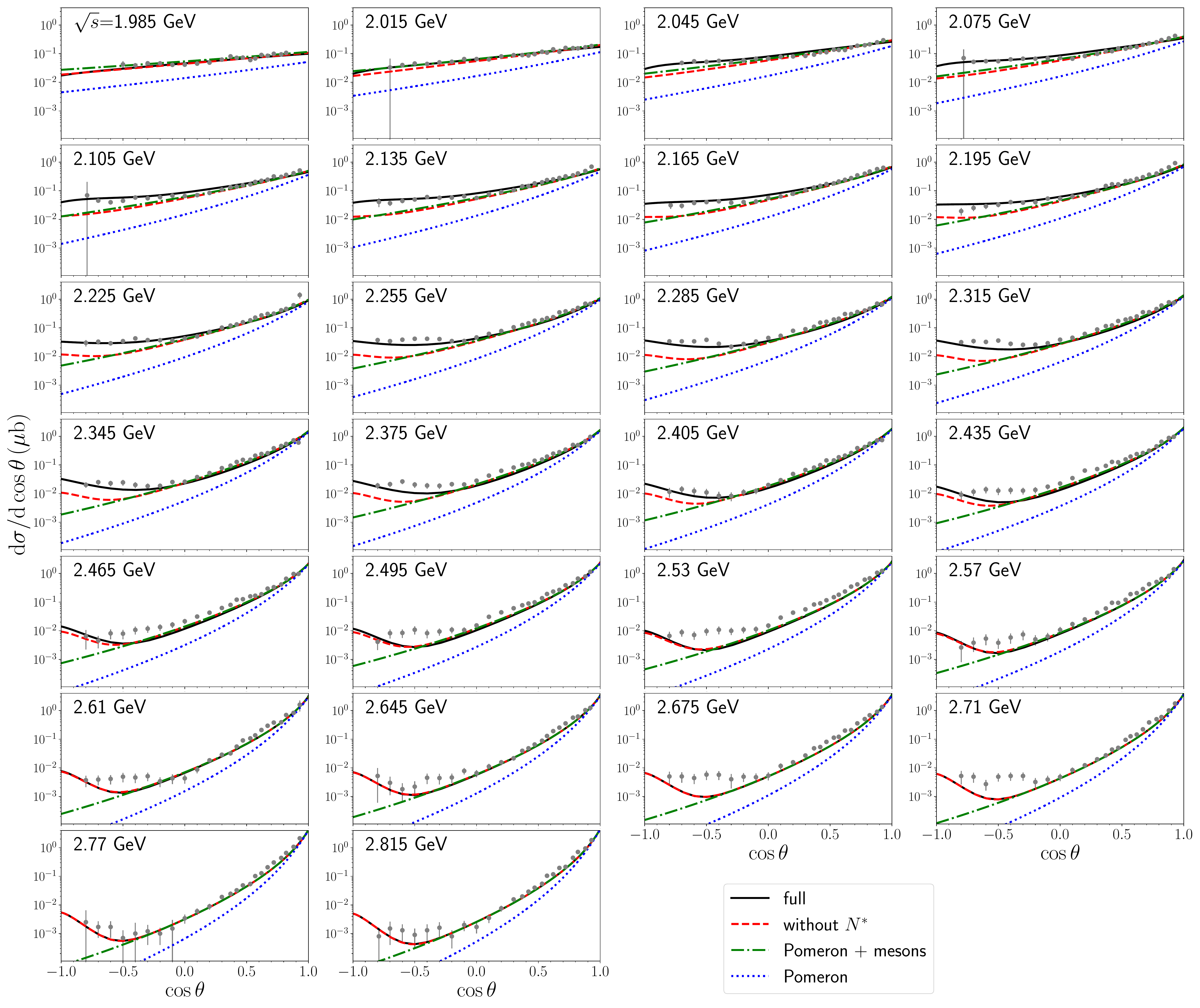}
\caption{Differential cross sections $d\sigma/d\cos\theta$ without FSI as functions of $\cos\theta$
for various c.m. energies $\sqrt s = W = (1.985 \mbox{--} 2.815)$~GeV.
The blue dotted lines are from the calculations including Pomeron exchange only.
The green dot-dashed lines show the sum of the Pomeron and meson exchange mechanisms.
The black solid lines and the red dashed lines represent the full contribution with and without $N^*$, 
respectively, which are almost identical at $\sqrt{s} \ge 2.57$~GeV.
The experimental data are from the CLAS Collaboration~\cite{CLAS14}.}
\label{fig:dcrst-data1}
\end{figure*}

\section{Results} \label{sec:results}

In this section, we present and discuss our numerical results for the cross sections of
$\gamma p \to \phi p$ and $\gamma \, \nuclide[4]{He} \to \phi \, \nuclide[4]{He}$.

\subsection{\boldmath $\gamma p \to \phi p$}

With the Pomeron-exchange model determined from the global fits to the total cross section data, 
as presented in Sec.~\ref{sec:impulse-A}, we first adjust the parameters of the meson-exchange and $N^*$
mechanisms to reproduce the CLAS data of Ref.~\cite{CLAS14}.
To simplify the fit of the $N^*$ parameters, we use the relevant information from the results of Ref.~\cite{KN19}.
The resulting parameters are shown and discussed in Sec.~\ref{sec:impulse-D}.

Our results on the differential cross sections of the $\gamma p \to \phi p$ reaction are presented in Fig.~\ref{fig:dcrst-data1}.
This shows that the full results (black solid lines) of our model could explain the differential cross section data 
of Ref.~\cite{CLAS14} very well.
The Pomeron exchange (blue dotted lines) accounts for far forward angle regions and begins to deviate from the data as $\cos\theta$ decreases.
The inclusion of various meson exchanges to the Pomeron exchange greatly enhances the results at $\cos\theta \geqslant 0$, as shown by green dot-dashed lines.
The main meson-exchange effects are due to the exchanges of scalar mesons, i.e., $a_0$ and $f_0$ mesons.
As shown by the red dashed lines, the direct $\phi$ radiation term is weak in the near threshold regions but becomes non-negligible as $\sqrt s$ ($=W$) increases at large
angles $\cos\theta \leqslant -0.5$.
To reproduce the CLAS data more accurately at $2.0 \leqslant \sqrt s \leqslant 2.4$ GeV and at large angles, we need additional ingredients.
We find that the $N^*$ excitations would play crucial roles in describing the data in these regions.

\begin{figure}[ht]
\centering
\includegraphics[width=\columnwidth]{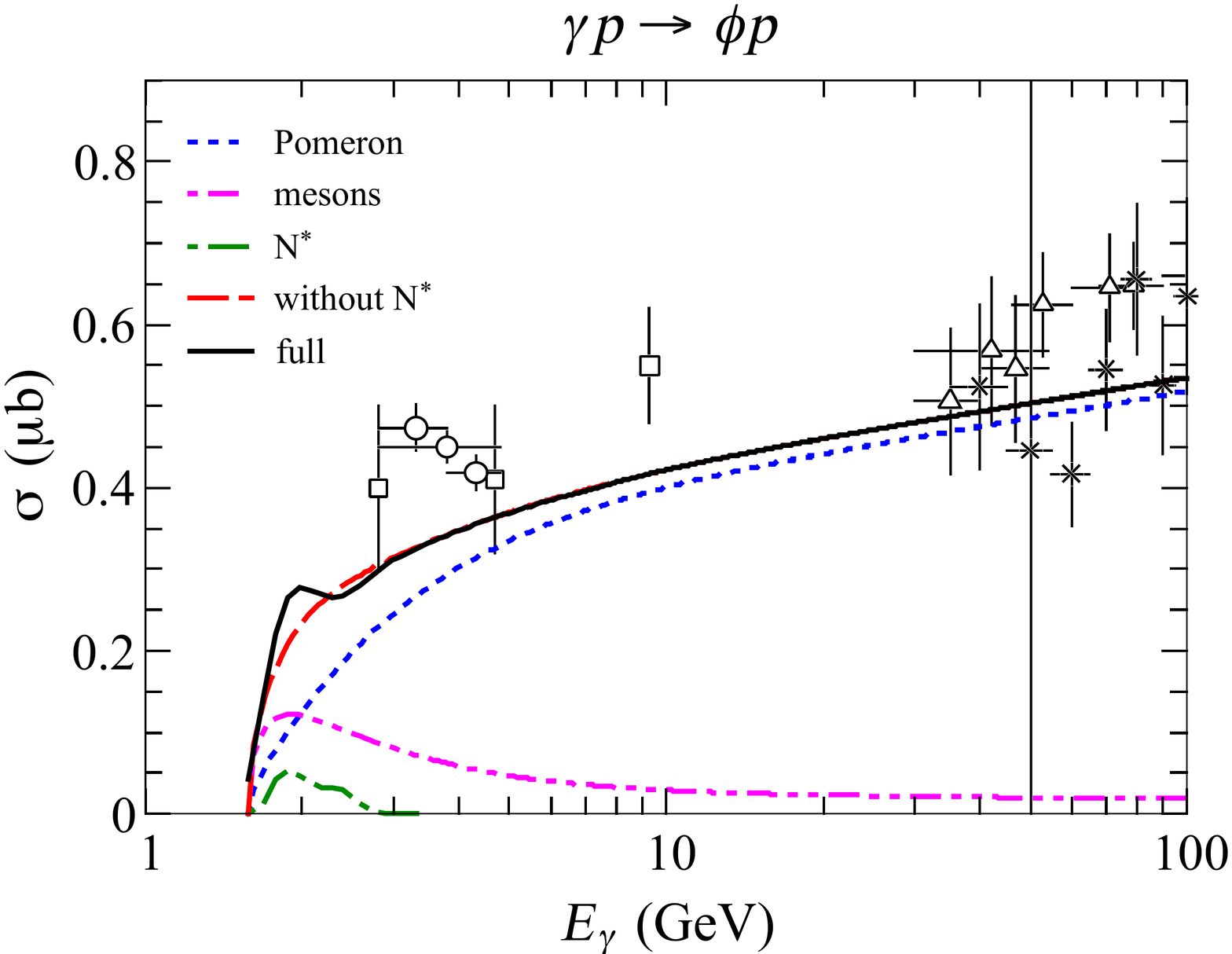}
\caption{Total cross section as a function of the photon energy in the laboratory frame $E_\gamma^{\rm lab}$.
The blue dotted line stands for the Pomeron exchange only.
The green dot-dashed and the magenta dot-dashed-dashed lines indicate the meson exchanges and $N^*$ contributions, respectively.
The black solid and the red dashed lines represent the full contributions with and without $N^*$, respectively. 
The experimental data are from Refs.~\cite{BCEK73} (open squares), \cite{BDLM82} (open circles), \cite{EDLM79} (open triangles),
and \cite{BOCG89} (stars).
}
\label{fig:phi-n-tcs}
\end{figure}

\begin{figure*}[t]
\centering
\includegraphics[width=\textwidth]{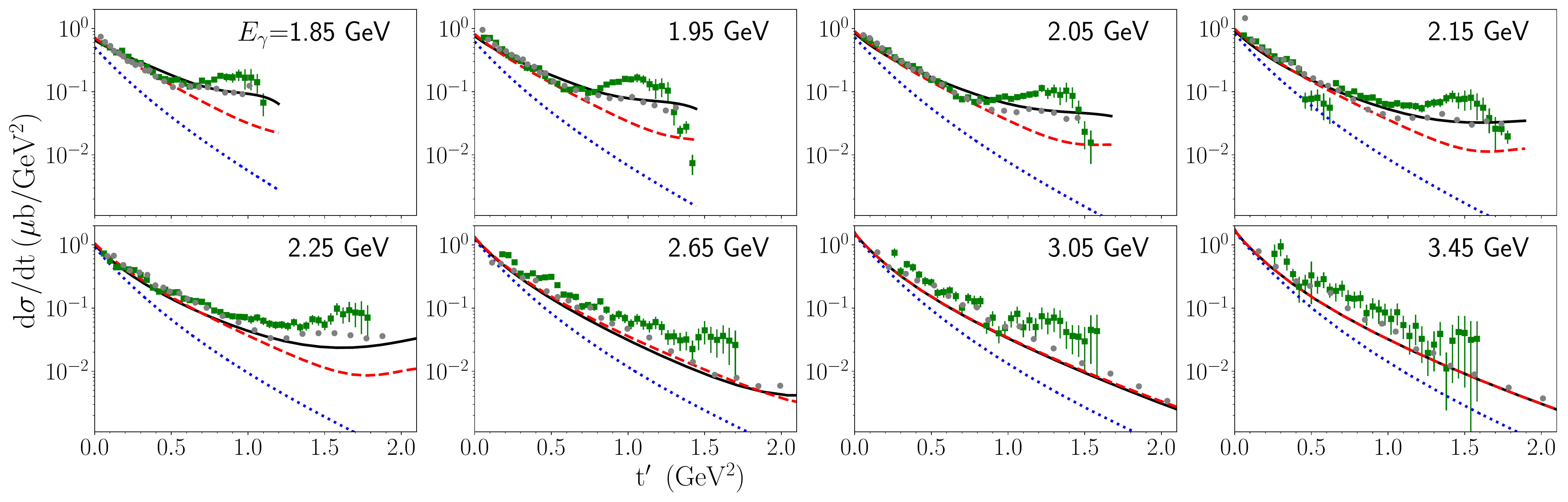}
\caption{Differential cross sections of $\gamma p \to \phi p$ without FSI at lower c.m. energies as functions of $t' = \vert t \vert -
\vert t \vert_{\rm min}$.
The blue dotted lines are from the calculations including Pomeron exchange only.
The black solid lines and the red dashed lines represent the full contribution with and without $N^*$, 
respectively.
The experimental data are from Ref.~\cite{CLAS14} (grey filled circles) and Ref.~\cite{CLAS13-c} (green filled squares).}
\label{fig:compare-data}
\end{figure*}

Figure~\ref{fig:phi-n-tcs} depicts the total cross section as a function of $E_\gamma$, the photon energy in the laboratory frame.
This shows that the Pomeron-exchange (blue dotted line) is clearly the dominant mechanism at 
high energy region, $E_\gamma > 10$~GeV.
However, in the near threshold region, the meson exchanges (magenta dot-dashed-dashed)
and $N^*$-excitations (green dot-dashed) contribute sizable effects.

There are a few comments on $\phi$ photoproduction in low energy region.
In Fig.~\ref{fig:compare-data}, we present the experimental data from Ref.~\cite{CLAS14} (grey filled circles) and from
Ref.~\cite{CLAS13-c} (green filled squares) at $E_\gamma \le 3.45$~GeV.
This manifestly shows the inconsistency between these two data sets.
In particular, the data of Ref.~\cite{CLAS13-c} have bump structures at large $t'$, while the structure is not seen or reduced in the
data of Ref.~\cite{CLAS14}. 
As our model parameters are determined based on the data of Ref.~\cite{CLAS14}, the bump structure cannot be simply reproduced 
by adjusting the meson-exchanges and $N^*$ parameters in the present approach.
Clearly, the differences between these two data sets need to be resolved before $N^*$ contributions can be more rigorously determined.

\subsection{\boldmath $\gamma p \to \phi p$ with FSI}

\begin{figure*}[t]
\centering
\includegraphics[width=\columnwidth]{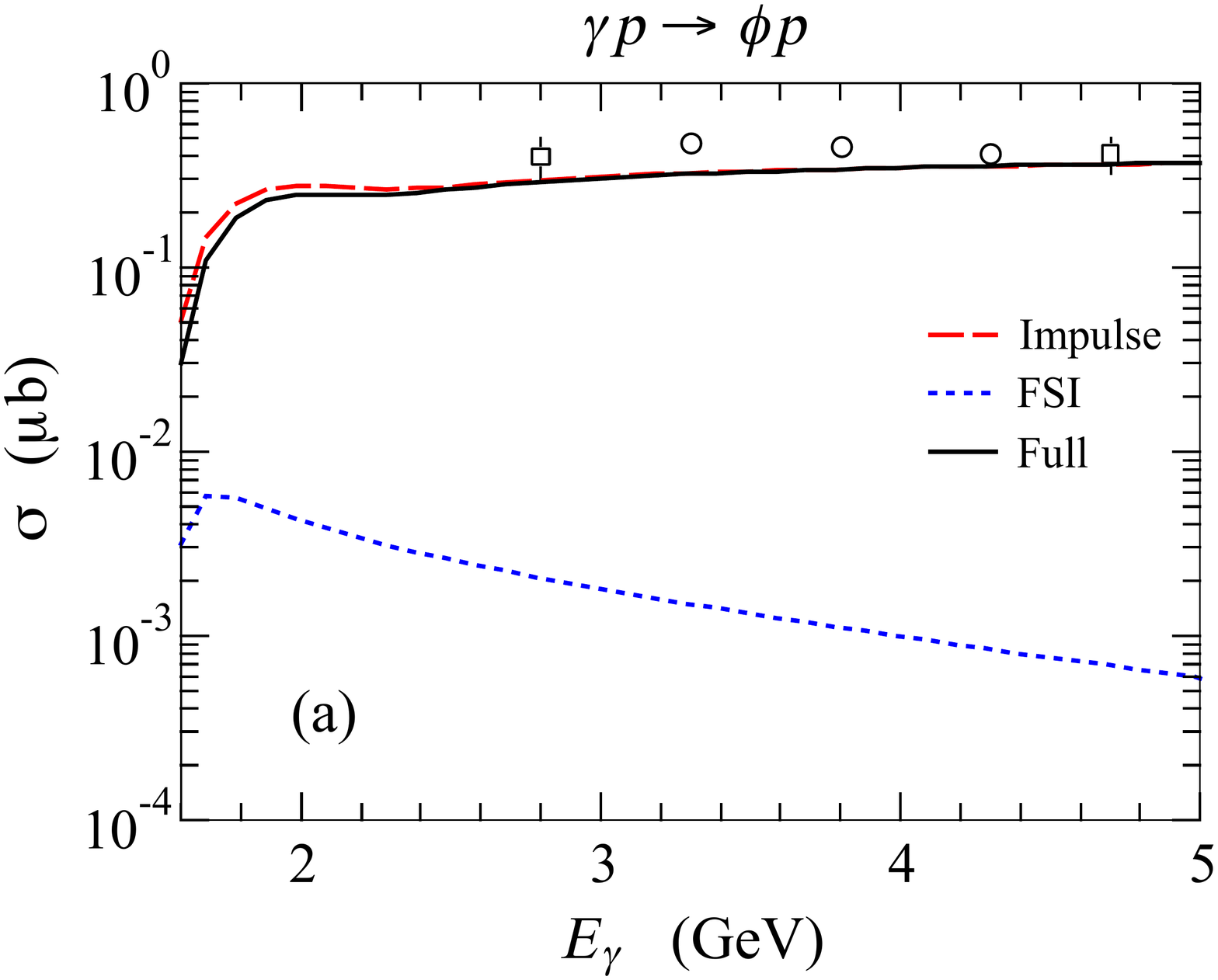} \hspace{0.5em}
\includegraphics[width=\columnwidth]{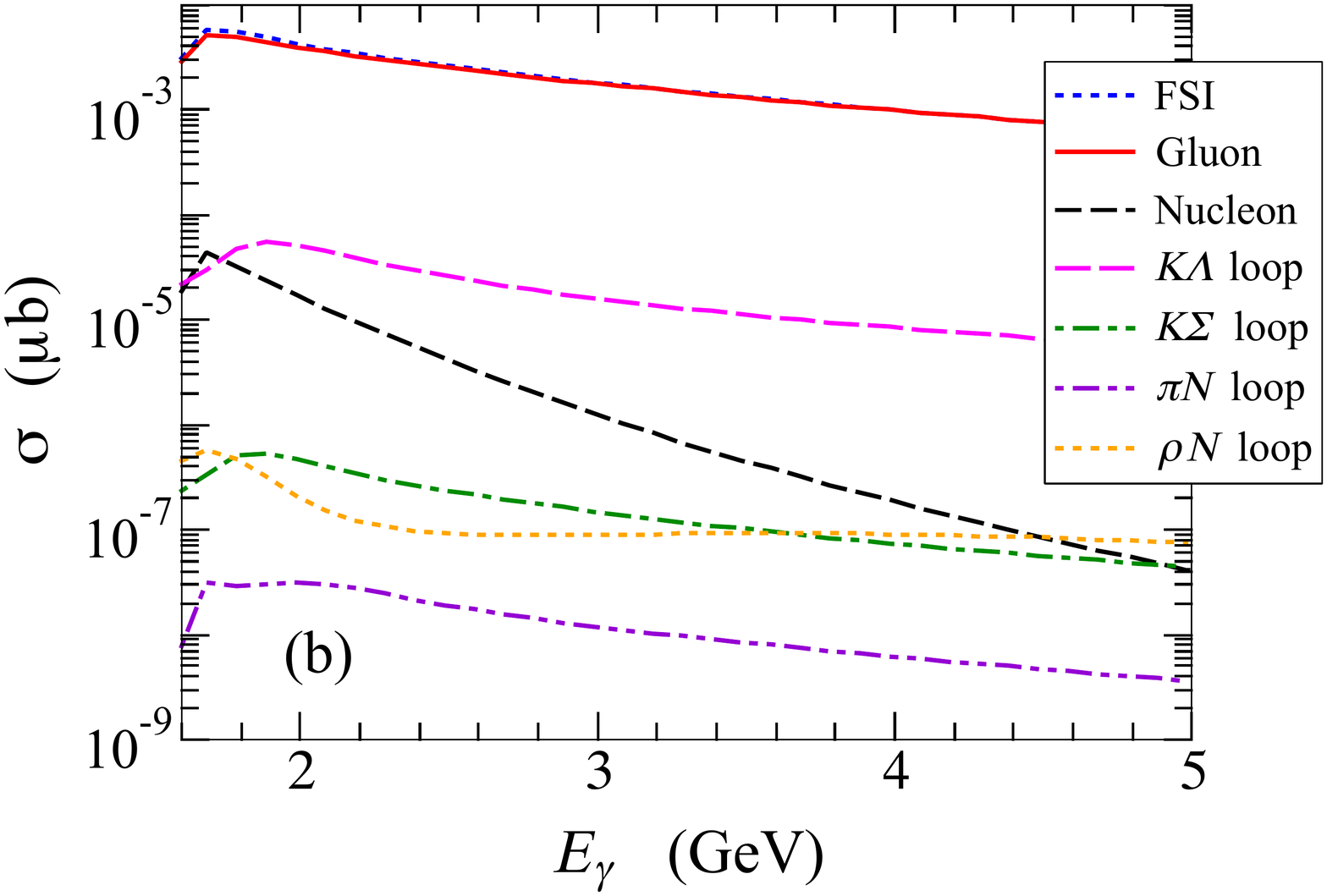}
\caption{Total cross sections of $\gamma p \to \phi p$ at low energies $E_\gamma$.
(a) The red dashed and the blue dotted lines stand for the impulse and the FSI
terms, respectively.
The solid black line indicates the full contribution.
The experimental data are from Refs.~\cite{BCEK73} (open squares) and \cite{BDLM82} (open circles).
(b) The individual contributions of the FSI terms.}
\label{fig:totcrst-fsi}
\end{figure*}

With the model of impulse approximation discussed in the previous subsection, we now consider the FSI in this subsection.
We present in Fig.~\ref{fig:totcrst-fsi}(a) the result of the total cross sections of $\gamma p \to \phi p$ with the FSI effects.
The impulse terms (red dashed lines) correspond to the full result in Fig.~\ref{fig:phi-n-tcs}.
The FSI effects represented by blue dotted lines are suppressed by factors of $10^2$--$10^3$ relative to the impulse terms
in the considered photon energy region.
The full results are given by the black solid lines, which are the sums of the impulse and FSI terms.
This reveals a destructive interference effect between them at $E_\gamma \leqslant 2.4$ GeV.

Figure~\ref{fig:totcrst-fsi}(b) displays the individual contributions of the FSI terms with the parameters 
determined in Sec.~\ref{sec:impulse}.
We find that the dominant contribution comes from the gluon-exchange interaction (red solid line)
which is more than two orders of magnitudes larger than the contributions of other FSI terms.
The $\phi N$ potential arising from the box-diagrams depends on the cutoff parameters in the form factors.
We choose $\Lambda_{MB}$ and $\Lambda_{MB}'$ in the range of the values used in the ANL-Osaka analysis 
of $\pi N$ and $\gamma N$ reactions~\cite{KLNS19}.
The contributions of the box-diagrams turn out to be rather weak.
The $K\Lambda$-loop diagram (magenta long dashed line) is the dominant among the considered four box diagrams.
The model parameters of the direct $\phi N$ coupling term (black short dashed line) [Figs.~\ref{fig:phin-mech}(b,c)] is 
determined from the $\phi$ radiation involved in the impulse term [Fig.~\ref{fig:gn-phin}(c,d)] using their similarities.
The direct $\phi N$ coupling contribution is comparable to the contribution of the $K\Lambda$-loop diagram near the threshold 
$E_\gamma \simeq 2$~GeV but falls off faster as $E_\gamma$ increases.

\begin{figure*}[t]
\centering
\includegraphics[width=\textwidth]{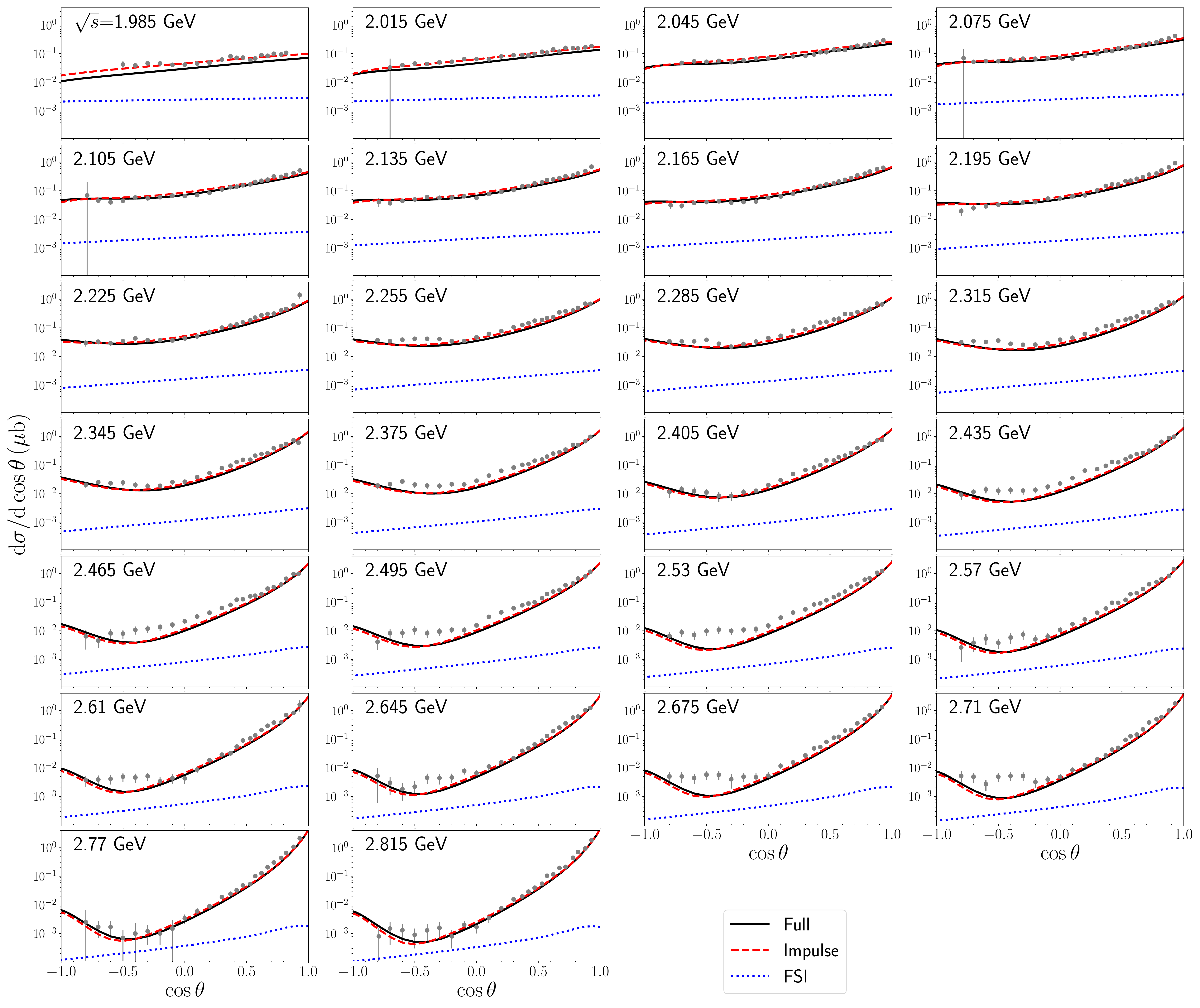}
\caption{Differential cross sections $d\sigma/d\cos\theta$ as functions of $\cos\theta$
for different photon energies $\sqrt s = (1.985 - 2.815)$ GeV with and without the FSI effects.
The red dashed and the blue dotted lines stand for the impulse and the FSI
terms, respectively.
The solid black lines indicate the full contributions.
The experimental data are from the CLAS Collaboration~\cite{CLAS14}.}
\label{fig:dcrst-fsi-1}
\end{figure*}

In Fig.~\ref{fig:dcrst-fsi-1}, we show the contributions from the FSI on differential cross sections $d\sigma/d\cos\theta$ 
as functions of $\cos\theta$.
The impulse model (red dashed lines) corresponds to the full results of Fig.~\ref{fig:dcrst-data1}.
This reveals that the FSI effects given by the blue dotted lines are clearly weak compared with the impulse terms over the
whole scattering angles.
Its shape is flat as a function of $\cos\theta$ near the threshold but becomes steeper as
$\sqrt s$ increases.

For the gluon-exchange potential $\mathcal{V}_{\rm gluon}=-v_0^{}  \exp(-\alpha r) /r$ of Eq.~(\ref{eq:yukawa}), we use
$\alpha = 500~$~MeV and $v_0 = 0.2$ which lie in the available ranges of $0.1 < v_0^{} < 1.0$ and
$0.3 < \alpha < 0.6 $~GeV.
If we use a larger value for $v_0$, the full results (black solid lines) begin to deviate from the data near the threshold because
the FSI term is desctuctive with the impulse term near the threshold region
as indicated in Fig.~\ref{fig:totcrst-fsi}(a).
Meanwhile, they interfere constructively at rather high energies and at backward
angles.

\subsection{\boldmath $\gamma \, \nuclide[4]{He} \to \phi \, \nuclide[4]{He}$}

\begin{figure}[t]
\centering
\includegraphics[width=\columnwidth]{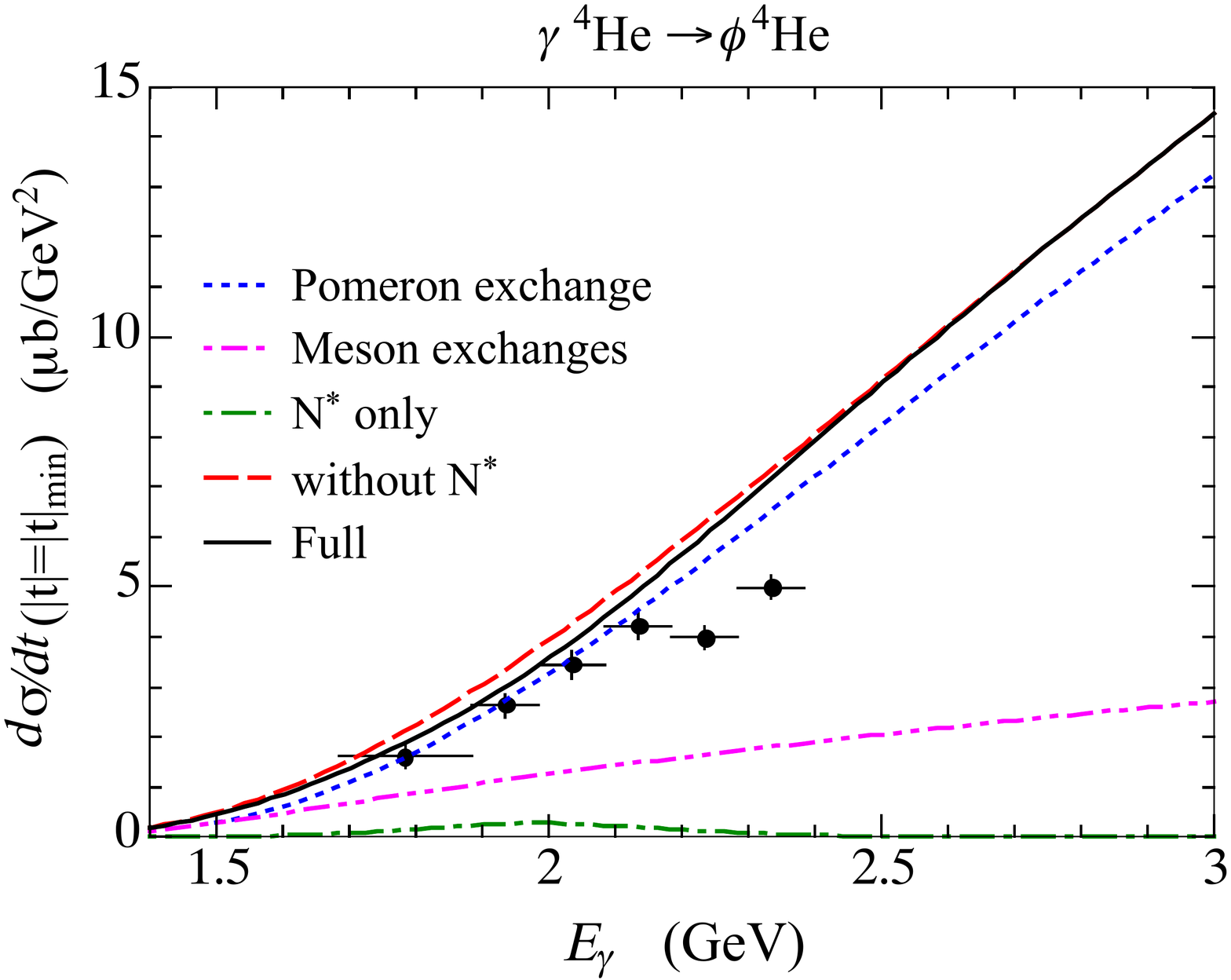}
\caption{Differential cross sections $d\sigma/dt$ at $t=t_{\mathrm{min}}$ as a function of $E_\gamma$ 
for the reaction of $\gamma \, \nuclide[4]{He} \to \phi \, \nuclide[4]{He}$.
The experimental data are from the LEPS Collaboration~\cite{LEPS17}.
}
\label{fig:totcrst-4he-1}
\end{figure}

In this subsection, we consider $\phi$ photoproduction from the \nuclide[4]{He} targets with the model for 
$\gamma p \to \phi p$ constructed in the present study.
We first work with the model without FSI and then present our results with FSI.
Presented in Fig.~\ref{fig:totcrst-4he-1} are our results  of differential cross sections $d\sigma/dt$ 
at $\vert t \vert =  \vert t \vert_{\rm min}$.
In this far forward region, the Pomeron-exchange contribution (dotted line) dominates, which can be expected.
If we keep only Pomeron-exchange and meson-exchange terms in our calculations, i.e., if we neglect the $N^*$ contributions,
we obtain the dashed line which somehow overestimates the experimental data~\cite{LEPS17}.
Although the $N^*$ contributions (dash-dotted line) are much weaker than the other mechanisms, inclusion of the $N^*$
results in a better description of the experimental data of LEPS Collaboration~\cite{LEPS17} in the region of
$E_\gamma < 2.2 $~GeV as shown by the solid line.
This observation ascribes to that the $N^*$ contributions interfere destructively with the other terms.
However, the structure shown by the experimental data at $E_\gamma \ge 2.2$~GeV cannot be explained by our model calculations
with variation of parameters.

\begin{figure}[t]
\centering
\includegraphics[width=\columnwidth]{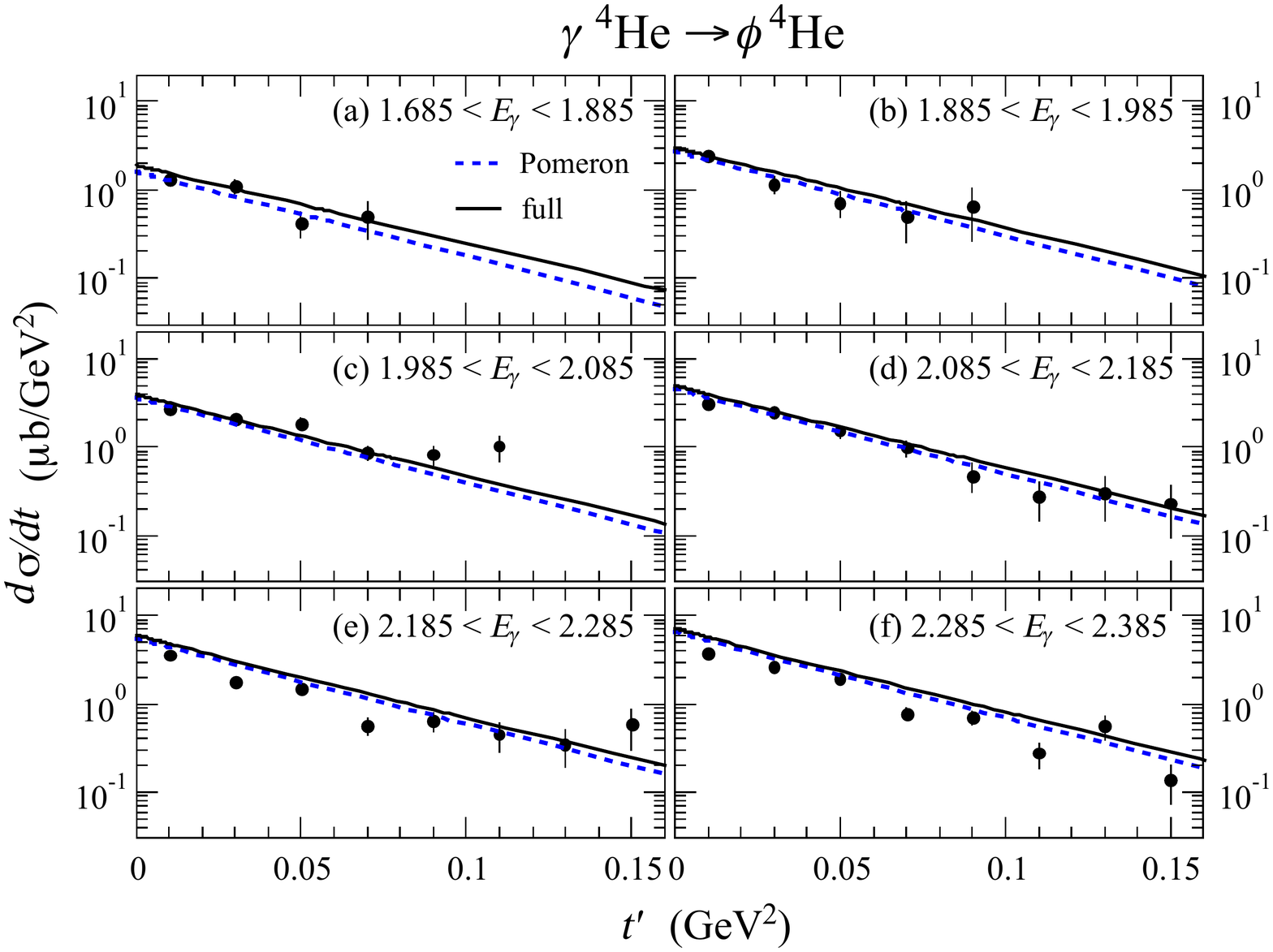}
\caption{Differential cross sections $d\sigma/dt$ of $\gamma\, \nuclide[4]{He} \to \phi\, \nuclide[4]{He}$
as functions of $t' \equiv \vert t \vert - \vert t \vert_{\rm min}$ at $1.685 < E_\gamma < 2.385$~GeV.
The dotted lines and the solid lines stand for the Pomeron and full contributions, respectively.
The experimental data are from Ref.~\cite{LEPS17}.}
\label{fig:dsdt-4he}
\end{figure}

\begin{figure}[t] 
\centering
\includegraphics[width=\columnwidth]{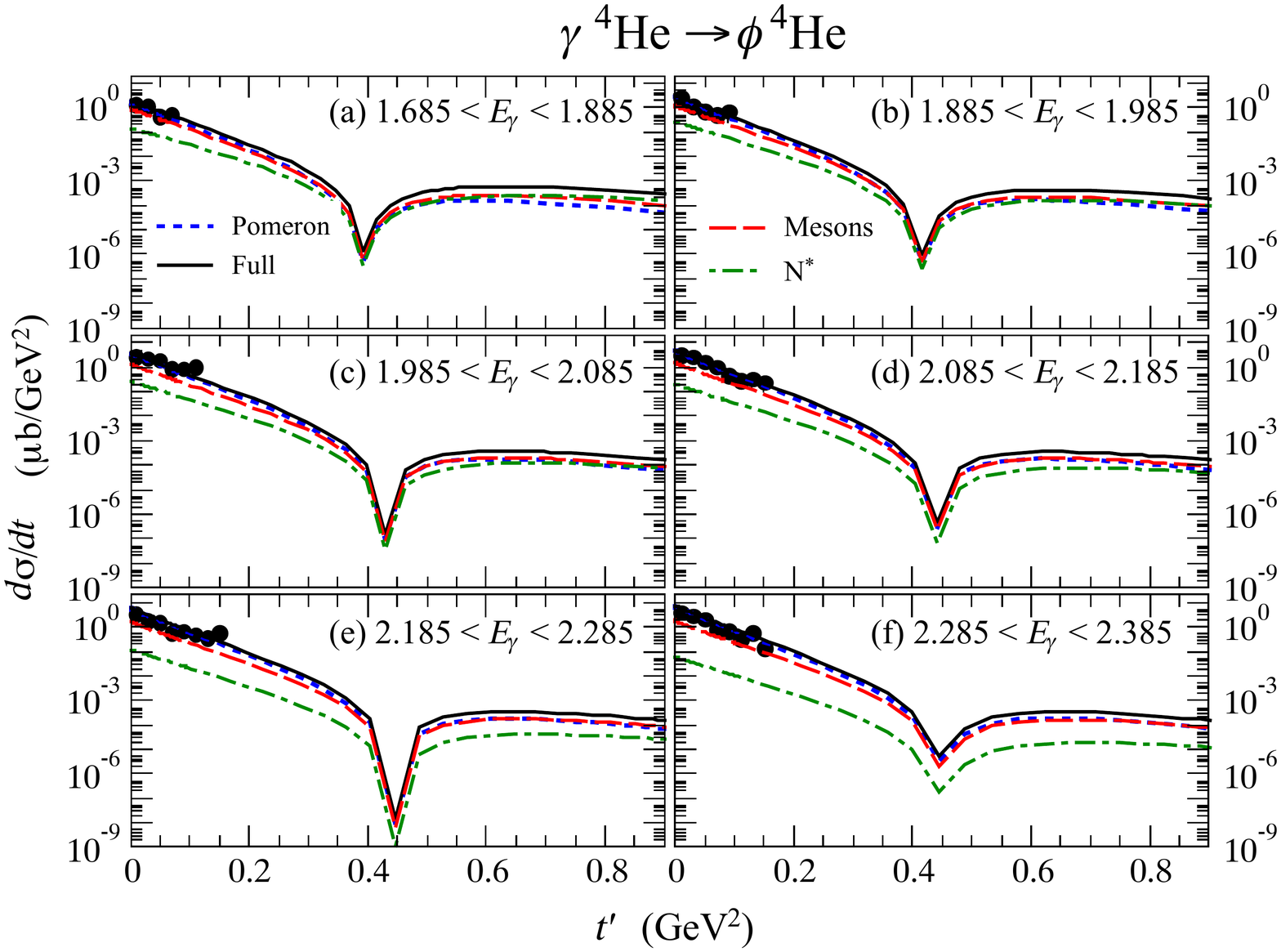}
\caption{Same as Fig.~\ref{fig:dsdt-4he} but for a wider range of $t' = \vert t \vert - \vert t \vert_{\rm min}$.
The dotted lines and the solid lines stand for the Pomeron and full contributions, respectively.
The dot-dashed lines are from the $N^*$ contributions.
The experimental data are from Ref.~\cite{LEPS17}.
}
\label{fig:dsdt-4he-1}
\end{figure}

In Fig.~\ref{fig:dsdt-4he}, we show differential cross sections $d\sigma/dt$ as functions of 
$t' \equiv \vert t \vert - \vert t \vert_{\rm min}$ for photon energies from 1.685 to 2.385~GeV.
We find that, in this forward angle region, the slopes of the dominant Pomeron exchange (dotted lines) 
match very well with the experimental data.
The calculated magnitudes of our full calculations (solid lines) are also in a reasonable agreement with the data, 
indicating that contributions from meson-exchanges and $N^*$ contributions are small in this far forward angle region. 
Their contributions become more sizable, however, at larger scattering angles as shown in Fig.~\ref{fig:dsdt-4he-1},
which also shows differential cross sections for a wider range of $t'$.
The bump structures shown at $t' \sim 0.4$~GeV$^2$ in Fig.~\ref{fig:dsdt-4he-1}
are due to the structure of the form factor $F_T(t)$ as seen 
in Eqs.~(\ref{eq:dsdt-imp-fsi}) and (\ref{eq:fsi}).
This structure could be tested in a future experiment.

\begin{figure}[t] \bigskip
\centering
\includegraphics[width=\columnwidth]{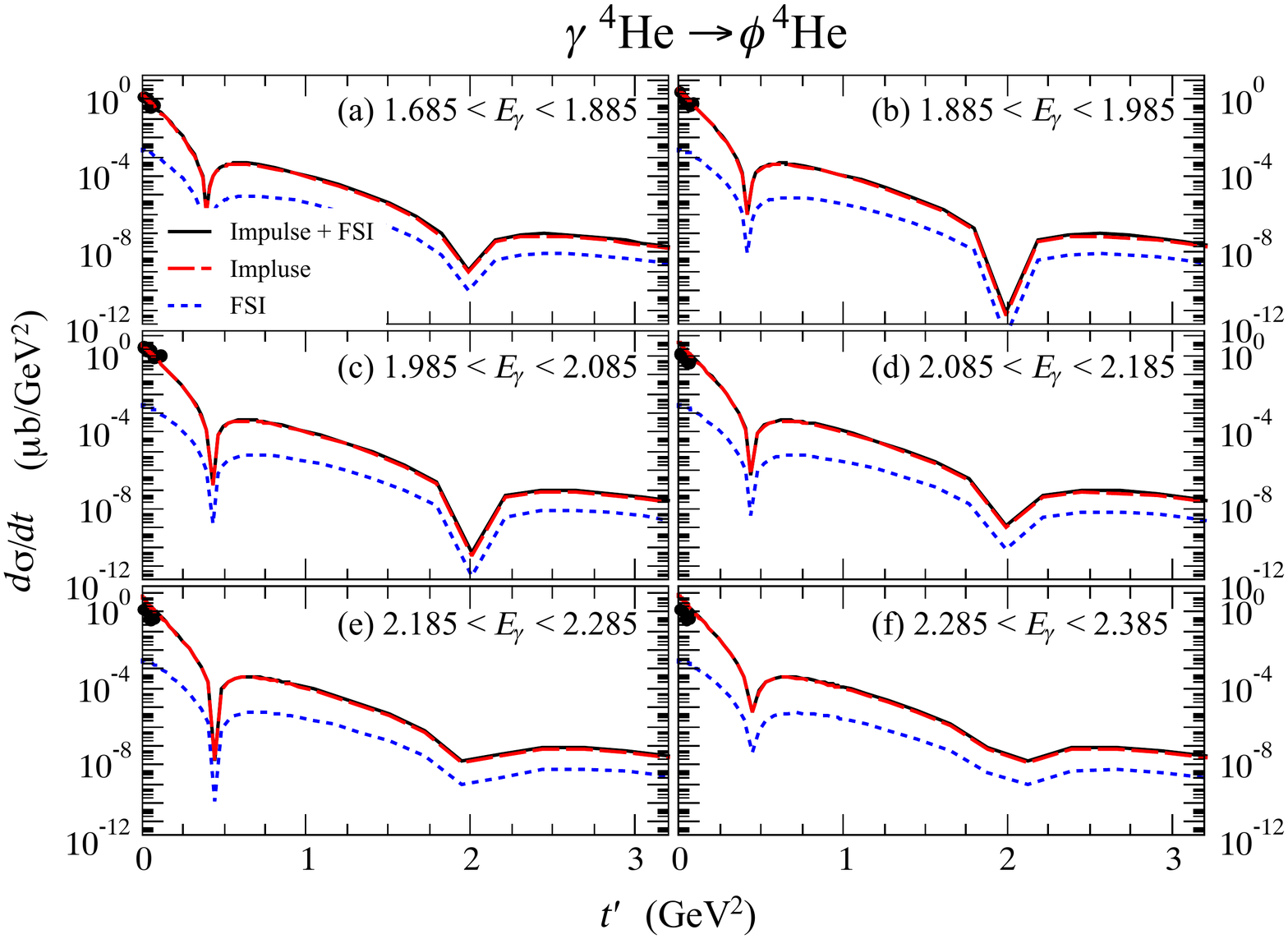}
\caption{Same as Fig.~\ref{fig:dsdt-4he-1} but for a wider range of $t' = \vert t \vert - \vert t \vert_{\rm min}$.
The dotted lines are the results with impulse approximation and FSI contributions are shown by the dashed lines.
Their sums are given by the solid lines.
The experimental data are from Ref.~\cite{LEPS17}.}
\label{fig:dsdt-4he-fsi}
\end{figure}

The results for $\phi$ photoproduction from the $\nuclide[4]{He}$ targets discussed so far are based on the
impulse approximation for $\gamma p \to \phi p$. 
We also carry out the calculations with FSI and our numerical results are presented in Fig.~\ref{fig:dsdt-4he-fsi}, 
which shows the results up to $t' = 3$~GeV$^2$.
Here, we can see that the FSI contributions (dotted lines) are very weak. 
This is not  surprising since the FSI effects are already very small for the production of the $\phi$ meson on the proton target 
as shown in Fig.~\ref{fig:dcrst-fsi-1}.
Furthermore, it is further reduced by the nuclear form factor $F(q)$ in the optical potential defined in Eq.~(\ref{eq:1st-opt}).

\section{Summary and Conclusion} \label{sec:summary}

In this work, we have investigated $\phi$ photoproduction on the nucleon targets and the \nuclide[4]{He} targets 
based on the approach developed by the ANL-Osaka collaboration group~\cite{KLNS19}.
The starting point is to construct a Hamiltonian which describes the mechanisms of $\phi$ photoproduction
 by the Pomeron-exchange, meson-exchanges, $\phi$ radiations, and $N^*$ excitation processes. 
 The final $\phi N$ interactions are described by the gluon-exchange and the direct $\phi N$ coupling 
 as well as the box-diagram arising from the $\pi N$, $\rho N$, $K\Lambda$, and $K\Sigma$ channels. 
The parameters of the Hamiltonian are determined by fitting the data of $\gamma p \rightarrow \phi p$~\cite{CLAS14}. 
The resulting Hamiltonian is then used to predict the coherent $\phi$ production on the \nuclide[4]{He} targets by using 
the distorted-wave impulse approximation.

For the proton target, we find that the Pomeron-exchange contributions are dominant in the very forward angles 
and the meson-exchange  mechanisms are crucial in obtaining a good fit to the experimental data in the 
large scattering angles, where  the $N^*$ excitations and $\phi$ radiation mechanisms are also required to
describe the data at the considered energy region. 
The final $\phi N$ rescattering effects, as required by the unitarity condition, are found to be very weak and the model
without the FSI effects is enough to reasonably describe the data.

For the \nuclide[4]{He} target, the calculated differential cross sections are in a very good agreement with the LEPS data~\cite{LEPS17} 
at low energies and for forward scattering angles. 
The FSI effects are also found to be negligible because of the further suppression originated from the nuclear form factor. 
The bump structures predicted by the present work could be tested by future experiments at a wider range of scattering angles.
However, the structure observed by the LEPS collaboration on $d \sigma/dt$ for far forward region
at $E_\gamma \sim 2.2 \mbox{--} 2.4$~GeV could not be explained by the present exploratory model and deserves further studies both in experiment and in
theory.

On the other hand, in the considered energy region, it would be interesting to investigate the role of other higher energy meson-baryon channels, 
such as  $\pi\Delta$, $\rho\Delta$, and $K\Lambda^*$, by extending the present work.
In addition, a full scale coupled-channels calculation as what was done by the ANL-Osaka Collaboration could be carried out
to go beyond the box-diagram approximations.
Our efforts in these directions will be reported elsewhere.

\acknowledgments
S.-H.K. and Y.O. are grateful to K. Tsushima for useful discussions at the early stage of
this work.
The work of S.-H.K. was supported by National Research Foundation (NRF) of Korea under
Grants No. NRF-2019R1C1C1005790 and No. NRF-2021R1A6A1A03043957.
S.i.N. was supported by NRF under Grants No. NRF-2018R1A5A1025563 and
No. NRF-2019R1A2C1005697.
T.-S.H.L. was supported by the Office of Science of the U.S. Department of Energy
under Contract No. DE-AC02-05CH1123.
The research of Y.O. was supported by Kyungpook National University Research Fund, 2021.


\end{document}